\documentclass[journal]{IEEEtran}
\usepackage{cite}
\usepackage{amsmath,amssymb,amsfonts}
\usepackage{algorithmic}
\usepackage{graphicx}
\usepackage{textcomp}
\usepackage[binary-units]{siunitx}
\usepackage{xcolor}
\usepackage{array}
\usepackage{float}
\newcolumntype{P}[1]{>{\centering\arraybackslash}p{#1}}
\usepackage{url}
\DeclareUnicodeCharacter{2212}{-}

\begin{document}

\title{3D homogenization of the T-A formulation for the analysis of coils with complex geometries}

\author{Carlos~Roberto~Vargas-Llanos,
        Felix~Huber,
        Nicol\`o~Riva,
        Min~Zhang,
        and~Francesco~Grilli% <-this % stops a space
%\thanks{{\review Funding acknowledgments.}}
%\thanks{Manuscript received April XX, XXXX; revised August XX, XXXX.}
\thanks{ C.R. Vargas-Llanos and F. Grilli are with the Institute for Technical Physics (ITEP) of the Karlsruhe Institute of Technology (KIT), Karlsruhe, Germany (e-mail: francesco.grilli@kit.edu). 
F. Huber and M. Zhang are with the department of Electronic and Electrical Engineering of the University of Strathclyde, Glasgow, United Kingdom.
N. Riva is with the Plasma Science and Fusion Center of the Massachusetts Institute of Technology (MIT), Cambridge, United States.}% <-this % stops a space
%\thanks{S. Lengsfeld is with Fraunhofer Institute for Energy Economics and Energy System Technology, 34119 Kassel, Germany}% <-this % stops a space
}

% The paper headers
\markboth{Journal of \LaTeX\ Class Files,~Vol.~XX, No.~X, Month~20XX}%
{Shell \MakeLowercase{\textit{et al.}}: Bare Demo of IEEEtran.cls for IEEE Journals}

\maketitle

% As a general rule, do not put math, special symbols or citations
% in the abstract or keywords.
\begin{abstract}
The modeling and analysis of superconducting coils is an essential task in the design stage of most devices based on high-temperature superconductors (HTS). These calculations allow verifying basic estimations and assumptions, proposing improvements, and computing quantities that are not easy to calculate with an analytical approach. For instance, the estimation of losses in HTS is fundamental during the design stage since losses can strongly influence the cooling system requirements and operating temperature. Typically, 2D finite element analysis is used to calculate AC losses in HTS, due to the lack of analytical solutions that can accurately represent complex operating conditions such as AC transport current and AC external applied magnetic field in coils. These 2D models are usually a representation of an infinitely long arrangement. Therefore, they cannot be used to analyze end effects and complex 3D configurations. In this publication, we use the homogenization of the T-A formulation in 3D for the analysis of superconducting coils with complex geometries where a 2D approach can not provide accurate analyses and verification of assumptions. The modeling methodology allows an easier implementation in commercial software (COMSOL Multiphysics) in comparison with the currently available 3D H homogenization, despite the complexity of the geometry. This methodology is first validated with a racetrack coil (benchmark case) by comparing the results with the well-established H formulation. Then, the electromagnetic behavior of coils with more complex geometries is analyzed. 
\end{abstract}
 
\begin{IEEEkeywords}
T-A formulation, 3D modeling, homogenization, high-temperature superconductors, AC losses, superconducting coil.
\end{IEEEkeywords}

\IEEEpeerreviewmaketitle

\section{Introduction}
\IEEEPARstart{T}{he} electrical properties of high-temperature superconductors (HTS) inspired several applications in different fields such as electrical machines~\cite{snitchler_10_2011},~\cite{dolisy_fabrication_2017},~\cite{frauenhofer_basic_2008}; fault current limiters~\cite{okakwu_application_2018}; magnets for scientific research~\cite{berrospe-juarez_estimation_2018};  energy storage~\cite{rong_developmental_2017} and transmission~\cite{fietz_high-current_2016}. The design of these devices usually require an electromagnetic analysis that allows verifying rated characteristics (i.e., power, torque, efficiency) as well as studying the behavior under different operating conditions. Moreover, losses in the superconducting tapes and wires must be estimated to design the cooling system. These AC losses can be decisive for the practical and economic realization of superconducting devices.

Several analytical solutions were developed to estimate losses in HTS tapes~\cite{brandt_type-ii-superconductor_1993},~\cite{norris_calculation_1970}, and infinite stacks of tapes~\cite{mawatari_critical_1996},~\cite{mawatari_critical_1997},~\cite{mikitik_analytical_2013}. However, these solutions are only valid under specific operating conditions such as AC transport current or applied uniform magnetic field. Therefore, they can not be directly used to estimate losses in most superconducting machines and equipment. For these reasons, a finite-element model is typically used to analyze the electromagnetic behavior and estimate losses in the HTS tapes. 

There are two popular formulations of Maxwell's equations that are commonly used to model superconductors by using the finite-elements method (FEM). The first one is based on the magnetic field strength ($\Vec{H}$) and was already used to study numerous applications~\cite{shen_review_2020},~\cite{shen_overview_2020}. The second one was introduced in~\cite{zhang_efficient_2016} and is based on the current vector potential ($\Vec{T}$) and magnetic vector potential ($\Vec{A}$). This T-A formulation is mostly used to analyze superconducting layers by applying a thin strip approximation. The approximation allows a reduction of dimensions that decreases the number of degrees of freedom and computation time. Therefore, it was used to study the cross-section of magnets~\cite{berrospe-juarez_estimation_2018} and electrical machines with hundreds and thousands of tapes~\cite{benkel_t-formulation_2020},~\cite{huang_fully_2020},~\cite{vargas-llanos_t-formulation_2020},~\cite{huber_t-formulation_2022}. 

Most of the FEM-based models used to study superconducting devices are 2D. They usually represent the cross-section of an infinite long or axisymmetric arrangement. Therefore, the end effects are not considered. Moreover, complex geometries and operating conditions such as saddle coils and twisted stacks of tapes under an external magnetic field can not be analyzed with a 2D model. For these reasons, several efforts were made for the development of tools and methodologies that allow 3D modeling of HTS coils~\cite{ando_development_2008},~\cite{zhao_establishment_2019},~\cite{gong_three-dimensional_2020},~\cite{zhang_dynamic_2020},~\cite{xu_3d_2021}. As part of these efforts, in 2014 the 3D H homogenization was introduced with the model of a racetrack coil~\cite{zermeno_3d_2014}. However, this approach requires the implementation of high resistivity layers in the homogenized domain. Five years later, the 3D homogenized T-A formulation was proposed by Huang et al.~\cite{huang_effective_2019}, who used it to calculate the AC transport losses of HTS racetrack coils. 

In this work, we expand the current knowledge by using the homogenization of the T-A formulation for the simulation of complex 3D HTS coils. The modeling approach is based on normal vectors defined by a local curvilinear coordinate system that considers the continuous shape and position of the tape and coil. Therefore, this general definition allows an easy implementation of the model despite the complexity of the geometry. The dependence of the critical current density on the magnetic flux density magnitude and direction can be taken into account by using the projection of the magnetic field in the direction of the unit vectors of the curvilinear system.

%{\review This general approach allows taking into account anisotropic material properties (such as $J_{\rm c}(\Vec{B})$) and easy implementation of the model, regardless of the complexity of its shape.}

%Therefore, this general definition allows an easy implementation of the model despite the complexity of the geometry. 

In this publication, we first describe the derivation of the governing equations in section~\ref{Section_T_A_Formulation}, with particular emphasis on the implementation of the normal vectors and how they can be defined in a commercial software (COMSOL Multiphysics). In the second place, we describe the HTS tape that was used in all the simulations (section~\ref{Section_Description_tape}). Then, we validate the modeling approach with a racetrack coil geometry that is considered a benchmark case (section~\ref{Section_racetrack_coil}), by comparing the results with the well-established  H  formulation~\cite{zermeno_3d_2014}. After the validation, the model and analysis of more complex geometries are introduced: saddle coil (section~\ref{Section_Saddle coil}), D-shape coil (section~\ref{Section_D-Shape coil}), and twisted coil (section~\ref{Section_Twisted coil}). Finally, the main conclusions of this work are summarized in section \ref{Conclu}. 

\section{T-A Formulation and Homogenization}\label{Section_T_A_Formulation}

In this section, we first present a small review of the T-A formulation and homogenization to set the basis that allows us to extend these concepts to 3D complex geometries. Then, the 3D T-A homogenization is explained based on unit vectors that are perpendicular to the flat face of the tape, and can be defined from the geometry of the coil.

\subsection{T-A formulation}
The first applications of the T-A formulation to study superconducting devices by using FEM were introduced in~\cite{zhang_efficient_2016},~\cite{liang_finite_2017}. These works presented the T-A formulation as an efficient approach to model HTS tapes with a high aspect ratio.  

This formulation of Maxwell's equations couples the current vector potential $\Vec{T}$ and magnetic vector potential $\Vec{A}$, which are defined by the current density $\Vec{J}$ and magnetic flux density $\Vec{B}$:

\begin{equation} 
\Vec{J}=\nabla \times \Vec{T} \label{T} 
\end{equation}

\begin{equation} 
\Vec{B}=\nabla \times \Vec{A}. \label{A}
\end{equation}

\begin{figure}%[hbt!]
\centerline{\includegraphics[width=0.5\textwidth]{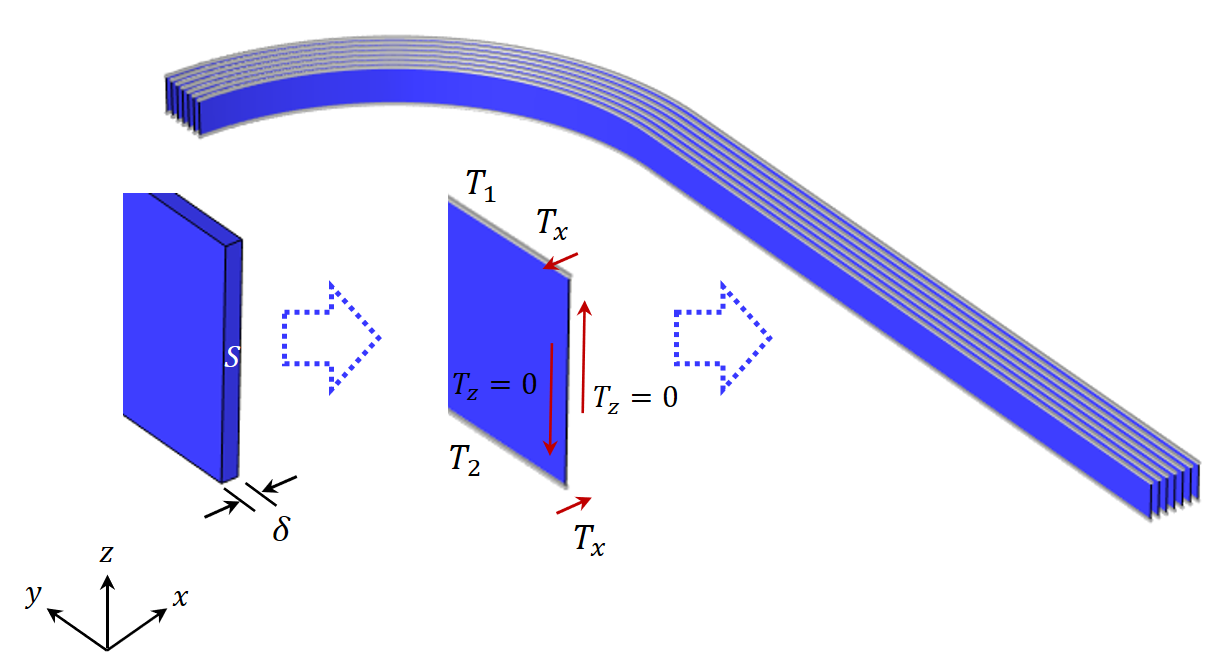} }
\caption{T-A formulation in 3D. $\Vec{T}$ is computed only in the superconducting domain (depicted in blue) while $\Vec{A}$ is computed everywhere. The tape's thickness is collapsed into a superconducting layer and the current is enforced by giving proper values of $T$ at the edges of the tape~\cite{berrospe-juarez_real-time_2019}.}
\label{T-A_formulation_3D}
\end{figure}

The magnetic vector potential is computed in all the domains under study by using Maxwell-Ampere's law ($\mu$ is the relative permeability of the material):

\begin{equation} \label{eq_Amp_A_formulation}
\nabla \times (\frac{1}{\mu} \nabla \times \Vec{A}) = \Vec{J}. \end{equation}

The current vector potential is calculated only in the superconducting domain by using Maxwell-Faraday's law:

\begin{equation} \label{eq_T_Faraday}
\nabla \times (\rho _{\rm HTS} \nabla \times \Vec{T}) = -\frac{\partial \Vec{B}}{\partial t}.
\end{equation}

The resistivity of the superconducting material is typically modeled with a power-law $\Vec{E}-\Vec{J}$ relation~\cite{rhyner_magnetic_1993}:

\begin{equation} \label{eq_HTS_resis}
\rho _{\rm HTS}=\frac{E_{\rm c}}{J_{\rm c}(\Vec{B})}  \Bigg| \frac{\Vec{J}}{J_{\rm c}(\Vec{B})} \Bigg| ^{n-1}.
\end{equation}

An approximation is done by considering that the superconducting layer in the tapes under study (for instance rare-earth barium copper oxide/REBCO tapes) has a very large width-to-thickness ratio. Therefore, we can collapse the thickness of the tape ($\delta$) as shown in figure~\ref{T-A_formulation_3D}. As a consequence, the current is able to flow only in a superconducting sheet and $\Vec{T}$ is always perpendicular to this sheet. For this reason, the current vector potential can be expressed as $T \cdot \Vec{n}$ ($\Vec{n}$ is a unit vector perpendicular to the superconducting layer)~\cite{zhang_efficient_2016}. 

%We can obtain the required boundary conditions at the edges of the tape/layer by integrating the current:
%The required boundary conditions at the edges of the tape can be obtained by integrating the current:
The transport current can be imposed by setting the boundary conditions for $T$ at the edges of the tape, as it is shown in equations (\ref{eq_enfo_I1}) and (\ref{eq_enfo_I2}):

\begin{equation} \label{eq_enfo_I1}
I=\iint_S \Vec{J} \cdot \,{\rm d}\Vec{s}=\iint_S (\nabla \times \Vec{T}) \cdot \,{\rm d}\Vec{s}=\oint_{\partial S} \Vec{T} \cdot \,{\rm d}\Vec{l}
\end{equation}

%As it can be observed in figure~\ref{T-A_formulation_3D}, equation (\ref{eq_enfo_I1}) can be reduced into: 

\begin{equation} \label{eq_enfo_I2}
I=(T_1-T_2)\delta.
\end{equation}

%which allows enforcing the current in coils by giving proper values of $T$ at the edges of each superconducting tape. 

In this publication, we use the order of the elements for discretization described in \cite{berrospe-juarez_real-time_2019} (linear elements for $\Vec{T}$ and quadratic elements for $\Vec{A}$) to avoid possible spurious oscillations.

\subsection{Homogenization in 3D}

\begin{figure}%[hbt!]
\centerline{\includegraphics[width=0.5\textwidth]{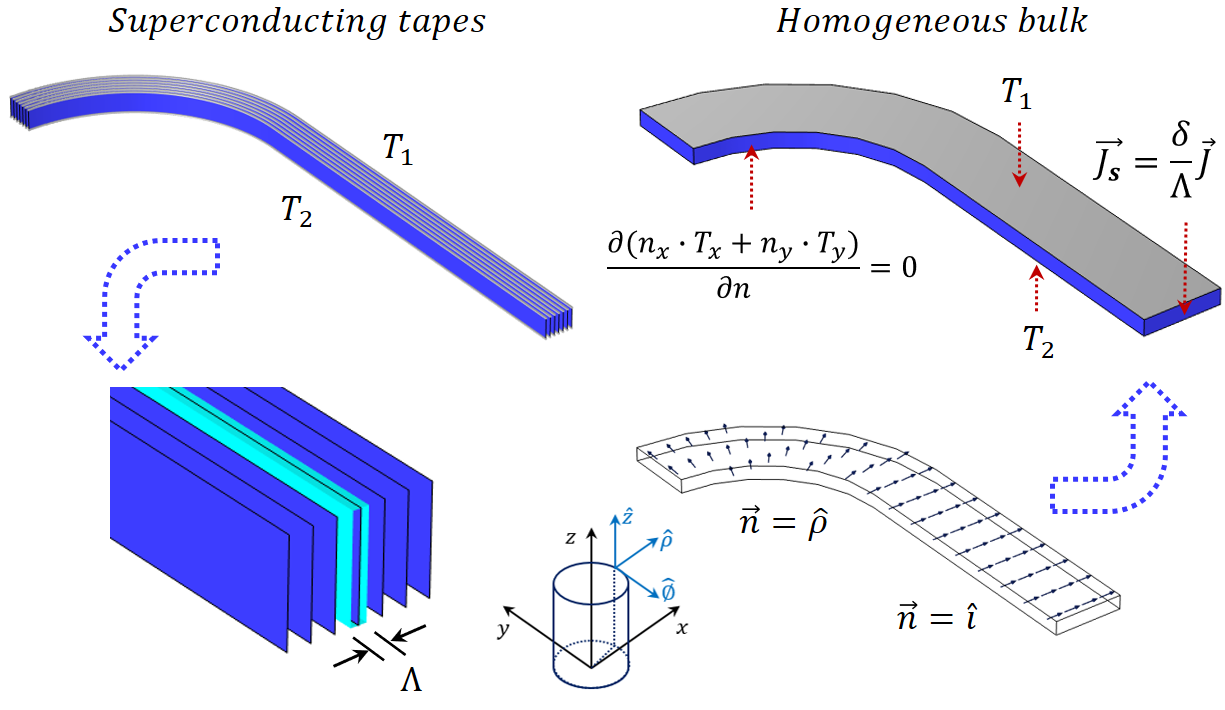} }
\caption{T-A homogenization in 3D. The superconducting sheets are replaced by a homogeneous block. The scale current density $\Vec{J}_s$ is introduced as a source term and boundary conditions $T_1$ and $T_2$ are applied to the upper and lower boundaries.}
\label{T-A_homogenization_3D}
\end{figure}

The homogenization technique assumes that the superconducting tapes that are wound in coils can be represented by an anisotropic bulk that can reproduce the overall electromagnetic behavior of the coil~\cite{zermeno_calculation_2013}. This allows reducing the number of degrees of freedom and the computation time. The technique was implemented by using the T-A formulation in 2D in~\cite{berrospe-juarez_real-time_2019},~\cite{berrospe-juarez_advanced_2021}. Therefore, we follow a similar approach to extend the homogenization into the 3D analysis of superconducting coils.

%~\cite{2020arXiv200602033B}

The procedure is summarized in figure~\ref{T-A_homogenization_3D}. We start from the arrangement of tapes modeled in the T-A 3D formulation as sheets and define a cell unit around them (with the same height of the superconducting tape and thickness $\Lambda$). The main characteristics of the tape will be impressed in this cell unit to transform the stack of tapes into a block. The scaled current in the homogenized bulk is defined for the $\Vec{A}$ calculation as:

\begin{equation} \label{eq_Js}
\Vec{J}_{s} = \frac{\delta}{\Lambda} \Vec{J}. 
\end{equation}

In principle, the homogenization makes use of Maxwell's equations in the same way as the 3D T-A formulation. Therefore, we keep the thin strip approximation of the tape that allows reducing the current vector potential into a scalar quantity. This means that even if the tapes are replaced by a homogeneous bulk, the current can only flow in the plane parallel to the original superconducting sheets. For example,  in the zoom presented in the bottom left corner of figure~\ref{T-A_homogenization_3D}, the current can only have $J_y$ and $J_z$ components and the current vector potential only has a $T_x$ component. The new homogenized block can be seen as a highly compressed group of superconducting tapes.
To represent a general geometry, we assume $\Vec{n}=\begin{bmatrix} n_x \\ n_y \\ n_z\end{bmatrix}$ and express equation (\ref{T}) as: 

\begin{equation} 
\begin{bmatrix} J_x \\ J_y \\ J_z\end{bmatrix}
=
\begin{bmatrix} 
\frac{\partial(T \cdot n_z)}{\partial y} - \frac{\partial(T \cdot n_y)}{\partial z} \\
\frac{\partial(T \cdot n_x)}{\partial z} - \frac{\partial(T \cdot n_z)}{\partial x} \\
\frac{\partial(T \cdot n_y)}{\partial x} - \frac{\partial(T \cdot n_x)}{\partial y}
\end{bmatrix}.
 \label{T2} 
\end{equation}

The magnetic vector potential is calculated in all the domains and we solve the current vector potential only in the superconducting domain by using Maxwell-Faraday's law:

\begin{equation} 
\begin{bmatrix} 
\frac{\partial(E_z)}{\partial y} - \frac{\partial(E_y)}{\partial z} \\
\frac{\partial(E_x)}{\partial z} - \frac{\partial(E_z)}{\partial x} \\
\frac{\partial(E_y)}{\partial x} - \frac{\partial(E_x)}{\partial y} 
\end{bmatrix} 
\cdot\Vec{n}
+
\begin{bmatrix} 
\frac{\partial(B_x)}{\partial t}\\
\frac{\partial(B_y)}{\partial t}\\
\frac{\partial(B_z)}{\partial t}
\end{bmatrix}
\cdot\Vec{n}
=0,
\label{eq_T_Faraday2} 
\end{equation}

{\noindent where $\Vec{n}$ can be easily determined in the T-A formulation because it is the vector perpendicular to the superconducting sheet. Therefore, it is usually defined by default in commercial software like COMSOL Multiphysics. However, once the stack of tapes is replaced, it is not easy to define $\Vec{n}$ inside the bulk. We have no longer a reference surface from which we can define the normal vector. A similar issue can be found in the 2D T-A homogenization when complex (rotated or curved) cross-sections are modeled in 2D. In this work, we present two alternatives to address this issue. First, we can analytically calculate $\Vec{n}$ by following the geometrical path of the tape before homogenization. If the tape is parallel to the $y-z$ plane then:}

\begin{equation} 
\Vec{n}=\hat i=\begin{bmatrix} 1 \\ 0 \\ 0\end{bmatrix}.
 \label{n_straight} 
\end{equation}

If the tape is wound in a circular shape with the center in the origin, then $\Vec{n}$ will be parallel to the radial vector in cylindrical coordinates $\hat \rho$:   
 
\begin{equation} 
\Vec{n}=\hat \rho=\begin{bmatrix} \frac{x}{\sqrt{x^2+y^2}} \\ \frac{y}{\sqrt{x^2+y^2}} \\ 0\end{bmatrix}.
 \label{n_circular} 
\end{equation} 

\begin{figure}[hbt!]
\centerline{\includegraphics[width=0.5\textwidth]{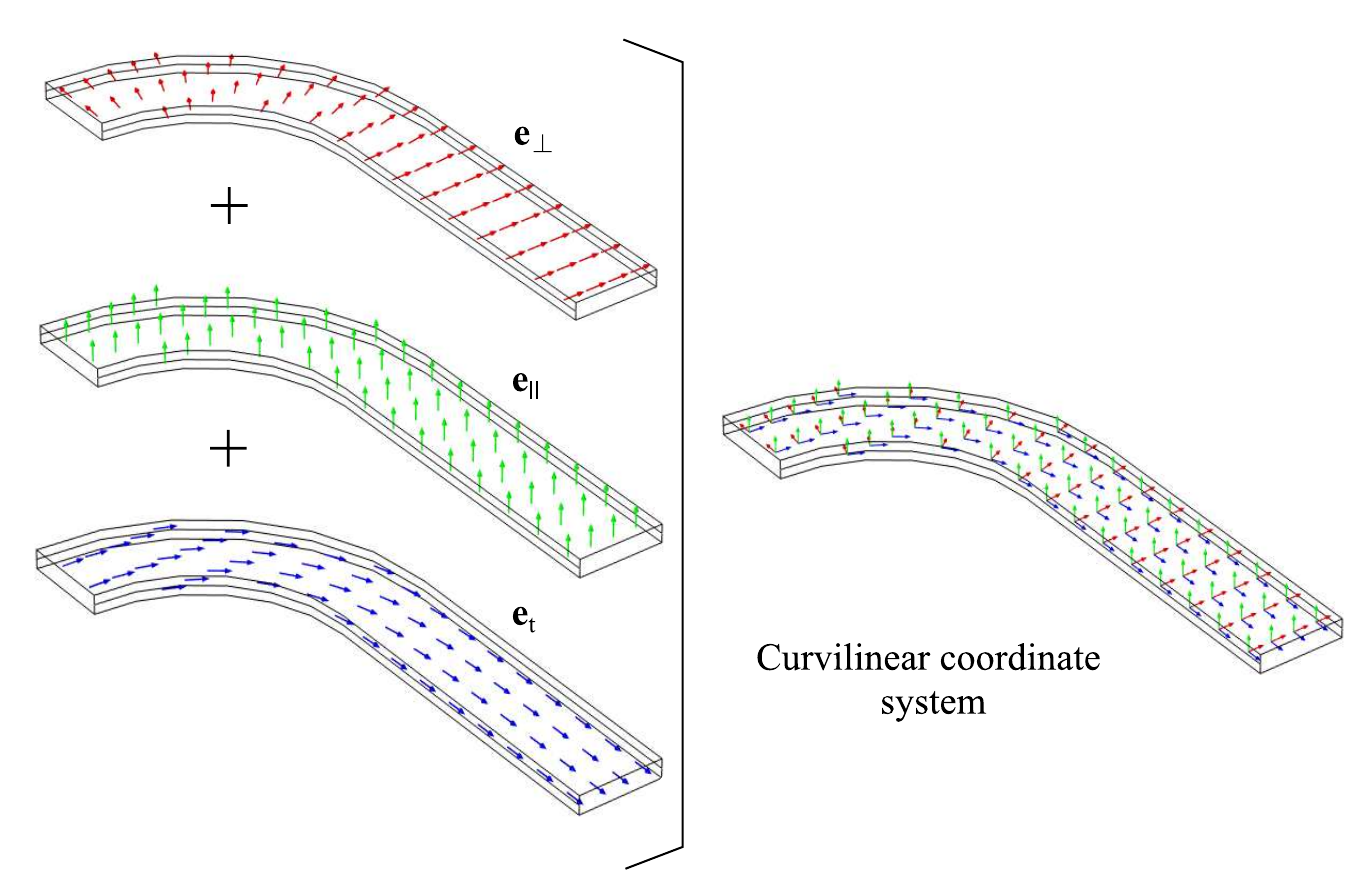} }
\caption{Curvilinear coordinate (CC) system example for the geometry shown in figure~\ref{T-A_homogenization_3D}. The CC system is defined based on the surface/boundaries of the 3D domain, which allows the definition of the base (unit) vectors of the system. These vectors can be used as references of the directions perpendicular to the wide face of the tape (${e}_{\bot}$), parallel to the wide face of the tape (${e}_{\parallel}$), and tangential to the winding direction (${e}_{t}$).}
\label{Figure_CCS}
\end{figure}

The normal vector can also be analytically defined by domains to represent more complex geometries as the one shown in the right corner of figure~\ref{T-A_homogenization_3D}. In this case, $\Vec{n}=\hat i$ in the straight section and $\Vec{n}=\hat \rho$ in the circular one. However, these analytic expressions can become very complex if we change the geometry of the coil, for example by introducing twisted or curved non-circular sections. Therefore, a second and more general solution is a curvilinear coordinate system defined in the superconducting domain. This approach creates a local coordinate system with curved lines that follow the shape of the superconducting domain~\cite{sjodin_using_nodate}, as it can be observed in figure~\ref{Figure_CCS} for the geometry described in figure~\ref{T-A_homogenization_3D}. The unit vectors of this local coordinate system can be defined for all possible geometries and used as normal/tangential vectors. This approach can be implemented in COMSOL Multiphysics by using the curvilinear coordinate module. 

Finally, we have to establish the boundary conditions to solve our problem. The homogenized bulk represents a densely packed group of HTS sheets. Each one of these sheets should transport the same current as its original counterpart~\cite{berrospe-juarez_real-time_2019}. Therefore, we apply Dirichlet boundary conditions to the upper and lower gray boundaries as expressed in equation (\ref{eq_enfo_I2}) and apply Neumann boundary conditions to the internal and external boundaries:

\begin{equation} 
\frac{\partial (n_x \centerdot T_x + n_y \centerdot T_y + n_z \centerdot T_z)}{\partial n}=0.
 \label{eq_Neumann_BC} 
\end{equation}

\section{Description of the HTS tape}\label{Section_Description_tape}

%%%%%%%%%%%%%%%%%%%%%%%%%%%%%%%%%%%%%%%%%%%%%%%%%%%%%%%%%%%%%%%%%%%%%%%%%%%%%%%%%%%%%%
\begin{figure}[hbt!]
\centerline{\includegraphics[width=0.5\textwidth]{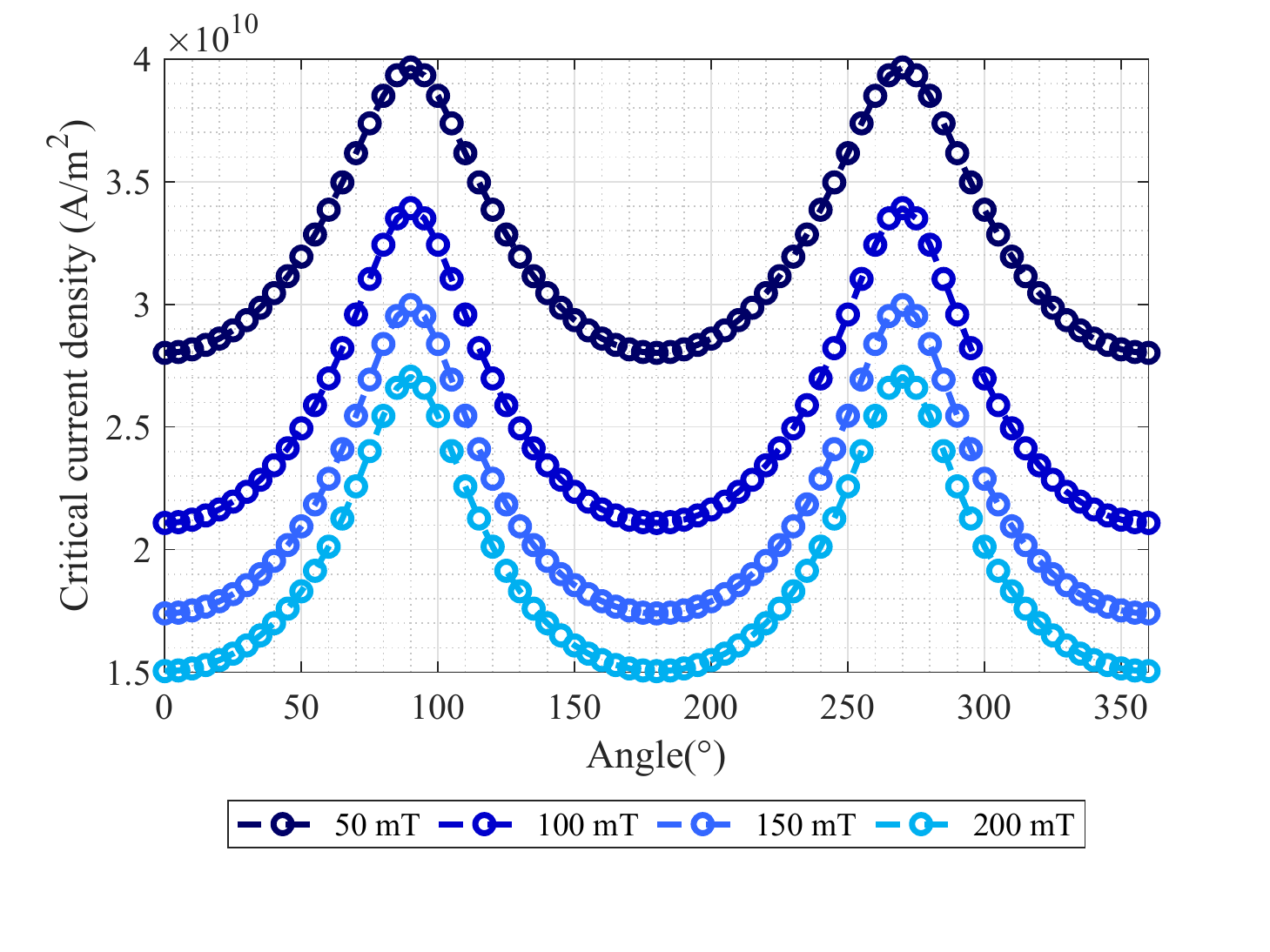} }
\caption{Critical current density behavior of the HTS tape for an external magnetic field magnitude of \SI{50}{\milli\tesla}, \SI{100}{\milli\tesla}, \SI{150}{\milli\tesla} and \SI{200}{\milli\tesla}. An angle of \ang{0} represents a field perpendicular to the wide face of the tape; an angle of \ang{90} represents a parallel one.}
\label{Jc_tape_plot}
\end{figure}
%%%%%%%%%%%%%%%%%%%%%%%%%%%%%%%%%%%%%%%%%%%%%%%%%%%%%%%%%%%%%%%%%%%%%%%%%%%%%%%%%%%%%%

The HTS tape used in all the simulations that we present in this publication is the one reported in~\cite{zermeno_3d_2014}. This allows direct comparison and validation of the modeling approach with the well-established 3D H homogenization. The critical current density dependence on the magnetic field amplitude and direction is described by the following equation:

\begin{equation} 
J_\textrm{c}(B_{\parallel},B_{\bot})=\frac{J_{c0}}{\left[1+\sqrt{(B_{\parallel}k)^2+B_{\bot}^2}/B_{c}\right]^b} ,
%\textbf{n}=\hat \rho=\begin{bmatrix} \frac{x}{\sqrt{x^2+y^2}} \\ \frac{y}{\sqrt{x^2+y^2}} \\ 0\end{bmatrix}.
 \label{JcB_equation} 
\end{equation} 

{\noindent where $B_{\parallel}$ and $B_{\bot}$ are the parallel and perpendicular components of the magnetic flux density, and the parameters $J_{c0}$, $k$, $B_c$ and $b$ have the following values \SI{49}{\giga\ampere\per\square\metre}, $0.275$, \SI{32.5}{\milli\tesla} and $0.6$~\cite{zermeno_3d_2014}. This is a tape with a critical current ($I_\textrm{c}$) of \SI{160}{\ampere} at \SI{77}{\kelvin} self-field. The behavior of the critical current density described in (\ref{JcB_equation}) can be better appreciated in figure~\ref{Jc_tape_plot}, where $J_\textrm{c}(B_{\parallel},B_{\bot})$ was plotted by considering four different magnetic field magnitudes with different directions.}

\section{Racetrack coil model and analysis}\label{Section_racetrack_coil}
%Critical current = 101.5 A - Methodology: The current was increased (ramp 100 A/s) and the change of the slop in the AC losses was detected at 101.5 A.
%Critical current AVG = 99.94 A & MAX = 98.26 A / P Methodology VMRZ

\begin{figure}[hbt!]
\centerline{\includegraphics[width=0.4\textwidth]{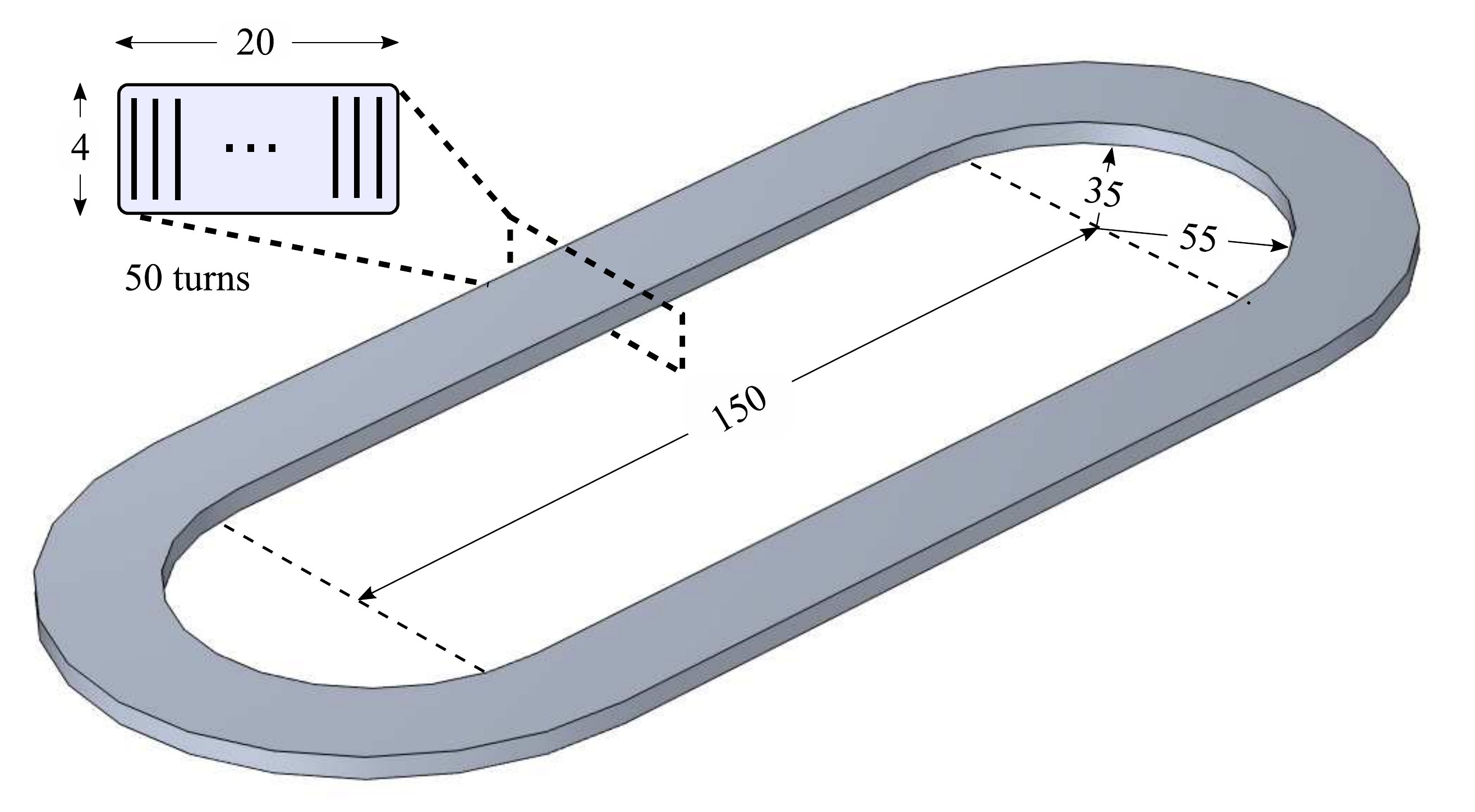} }
\caption{Geometry and dimensions of the racetrack coil under study~\cite{zermeno_3d_2014}. The coil is made with 50 turns of HTS tape, which creates a stack \SI{4}{\milli\meter} wide and \SI{20}{\milli\meter} height in the cross-section of the coil. All the dimensions are in millimeters.}
\label{Racetrack_geom}
\end{figure}

In this section, we study the racetrack coil presented in~\cite{zermeno_3d_2014}. This kind of coil is used for example in the rotor of superconducting electrical machines for wind turbine applications~\cite{abrahamsen_large_2012}. They have two straight and two round parts as shown in figure~\ref{Racetrack_geom}, where the dimensions of the racetrack coil under analysis are also given. The critical current of the coil was estimated by using the methodology described in~\cite{zermeno_self-consistent_2015} in a 2D model that represents the cross-section of the middle of the straight part of the coil. According to this approach and the average criterion, the critical current of the coil is \SI{100}{\ampere}.  

The normal vector ($\Vec{n}$) can be defined in this case by domain by following the original path of the superconducting tape or by using a curvilinear coordinate system, as it was mentioned in section \ref{Section_T_A_Formulation}. We used both approaches in this case and obtained identical results. For this reason, we only present in this publication the results of the model that uses the curvilinear coordinate system.

%\begin{table}[hbt!]
%\centering
%\caption{Dimensions of the racetrack coil.}
%\label{table_dimensions_racetrack}
%\begin{tabular}{l c} 
%\hline			
%Internal radius ($r_1$)                     &	\SI{35}{\milli\meter}	\\
%External radius ($r_2$)                     &	\SI{55}{\milli\meter}	\\
%Straight part length ($2 \centerdot y_0$)    &	\SI{150}{\milli\meter}	\\
%Thickness                                    &	\SI{4}{\milli\meter}	\\
%Turns                                        &	50	\\
%\hline
%\end{tabular}
%\end{table}

%%%%%%%%%%%%%%%%%%%%%%%%%%%%%%%%%%%%%%%%%%%%
\begin{figure*}[hbt!]
  \centering
  \includegraphics[width=0.9\textwidth]{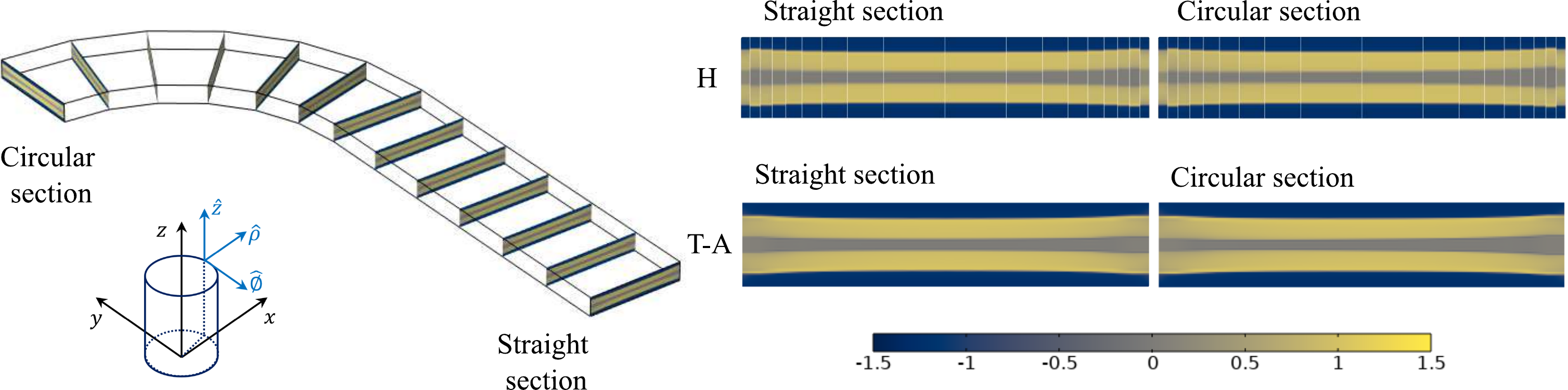}
  \caption{Normalized current density ($J/J_{\textrm{c}}$) in one-fourth of the racetrack coil for AC transport current ($f=\SI{50}{\hertz}$ and $I_{\rm peak}=\SI{100}{\ampere}$) when the current is equal to zero, after the first half period of the sinusoidal cycle. A zoom in the middle of the circular section and the straight section of the racetrack coil is shown on the right, where a comparison between the 3D H and T-A homogenization results is presented.}
  \label{Racetrack_J_Jc}
\end{figure*}
%%%%%%%%%%%%%%%%%%%%%%%%%%%%%%%%%%%%%%%%%%%%%

\begin{figure}[hbt!]
\centerline{\includegraphics[width=0.5\textwidth]{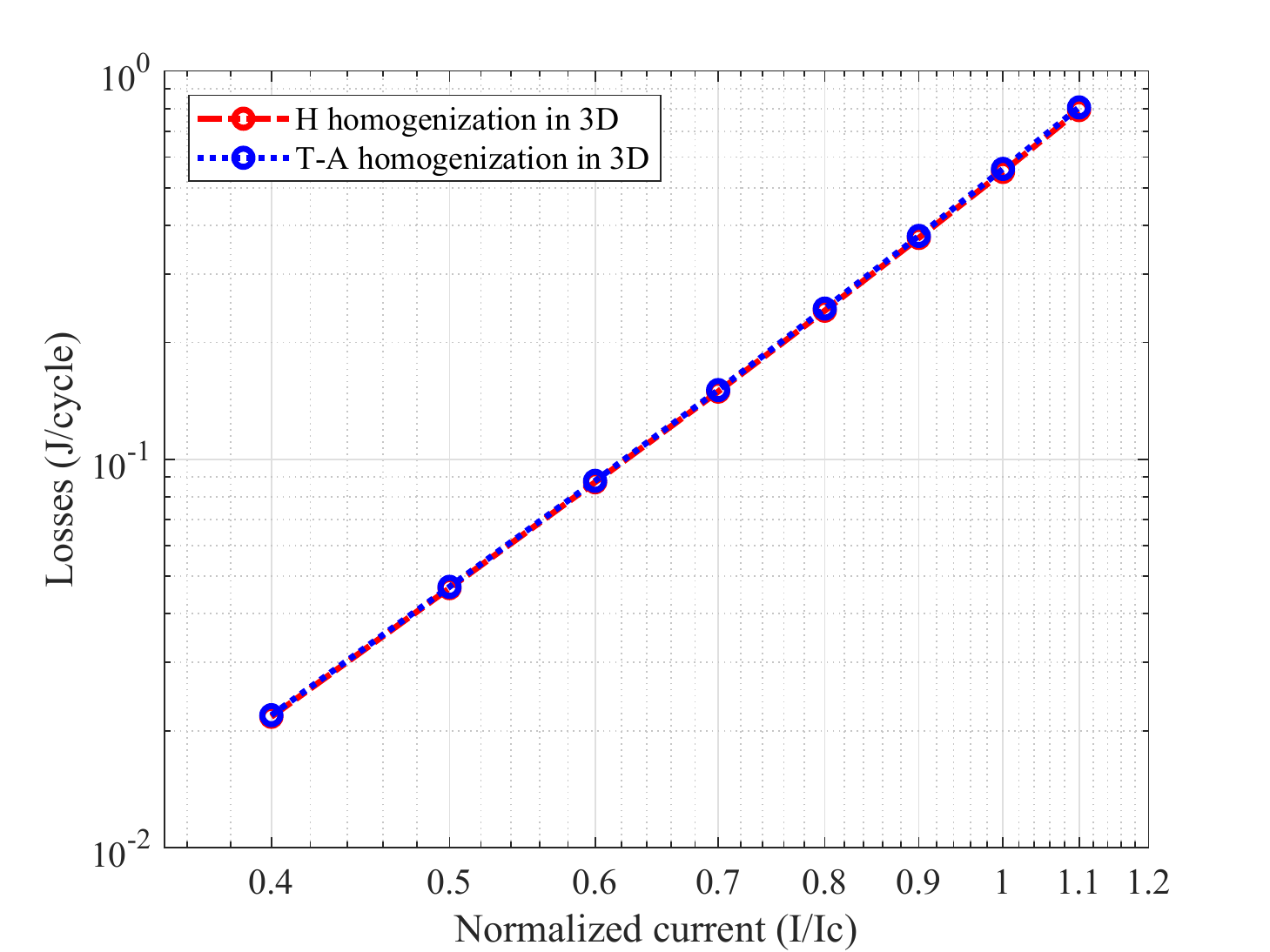} }
\caption{Estimation of losses in the racetrack coil due to AC transport current ($f=\SI{50}{\hertz}$)  as a function of the current amplitude by using the T-A 3D homogenization and the H 3D homogenization.}
\label{Racetrack_coil_losses}
\end{figure}

We assume in this section the same parameters presented in~\cite{zermeno_3d_2014} to model the superconducting tape. Similar to what was done in~\cite{zermeno_3d_2014}, we modeled only one-eighth of the coil by taking advantage of the symmetries.

The critical current density dependence on the magnetic field magnitude and direction was modeled by using the parallel ($B_{\parallel}$) and perpendicular ($B_{\bot}$) components of the magnetic flux density~\cite{zermeno_3d_2014}. For this reason, these components were also calculated by domain. In the circular section, they were computed as $B_{\parallel}=B_z$ and  $B_{\bot}=B_\rho$, by assuming that the circular region is centered at the origin. In the straight section, they were computed as $B_{\parallel}=B_z$ and  $B_{\bot}=B_x$, by considering that the straight region is parallel to the $y$ axis.

%We can also observe in this figure a comparison of the normalized current density distribution obtained with the 3D H homogenization and 3D T-A homogenization

The normalized current density in the cross-sections of the racetrack coil for AC transport current ($f=\SI{50}{\hertz}$ and $I_{\rm peak}=\SI{100}{\ampere}$) without external magnetic field, when the current is equal to zero (after the first half period of the sinusoidal cycle), is shown in figure~\ref{Racetrack_J_Jc}. In this figure we can also observe a comparison of the normalized current density distribution obtained with the 3D H homogenization and 3D T-A homogenization, at the middle of the straight and circular sections. The current density distribution obtained by using the 3D T-A homogenization is in good agreement with the one computed with the 3D H homogenization, and it is similar along the coil. We only observed a small difference between the middle of the straight and circular sections when the current is close to its maximum value, as it was reported in~\cite{zermeno_3d_2014}. Figure~\ref{Racetrack_coil_losses} shows the losses due to AC transport current estimated with the 3D T-A homogenization model. The results are in good agreement with the 3D H homogenization calculation for all the studied current values, with a maximum relative error lower than \SI{2}{\percent}. 

%At first sight, it seems that the current density distribution is very similar inside the coil. However, if we make a zoom in the middle of the circular and the straight section of the coil when we are close to the maximum current (positive or negative), we can notice that the innermost part of the circular section fills first with the critical current. This behavior can be related to a higher magnetic flux density in this internal area of the coil, in comparison with the straight region. Therefore, the analysis of the behavior of this section can be decisive in the operation of the coil. 

The main advantage of the 3D T-A homogenization is that it is relatively easy to implement. In particular, it does not require 2D integral constraints or high resistivity zones to prevent current sharing between subdomains of different groups of tapes. This restriction, which is quite tricky to implement in the 3D H homogenization~\cite{zermeno_3d_2014}, is included in the essence of the T-A homogenization by applying the thin strip approximation which constraints the current to flow in the plane parallel to the tape and allows reducing the current vector potential into a scalar quantity. Moreover, the proposed normal vector approach (based on a curvilinear coordinate system definition) allows simulating any geometrical arrangement. Therefore, it represents an effective and practical tool for the analysis of complex geometries.

\section{Saddle coil model and analysis}\label{Section_Saddle coil}
%Critical current = 101.5 A - Methodology: The current was increased (ramp 100 A/s) and the change of the slop in the AC losses was detected at 101.5 A.
%Critical current AVG = 99.75 A & MAX = 98.17 A / P Methodology VMRZ

Since the modeling technique was already validated in the previous section, we can proceed with the analysis of more complicated geometries that are currently being investigated for the development of superconducting devices. 

\begin{figure}[hbt!]
\centerline{\includegraphics[width=0.5\textwidth]{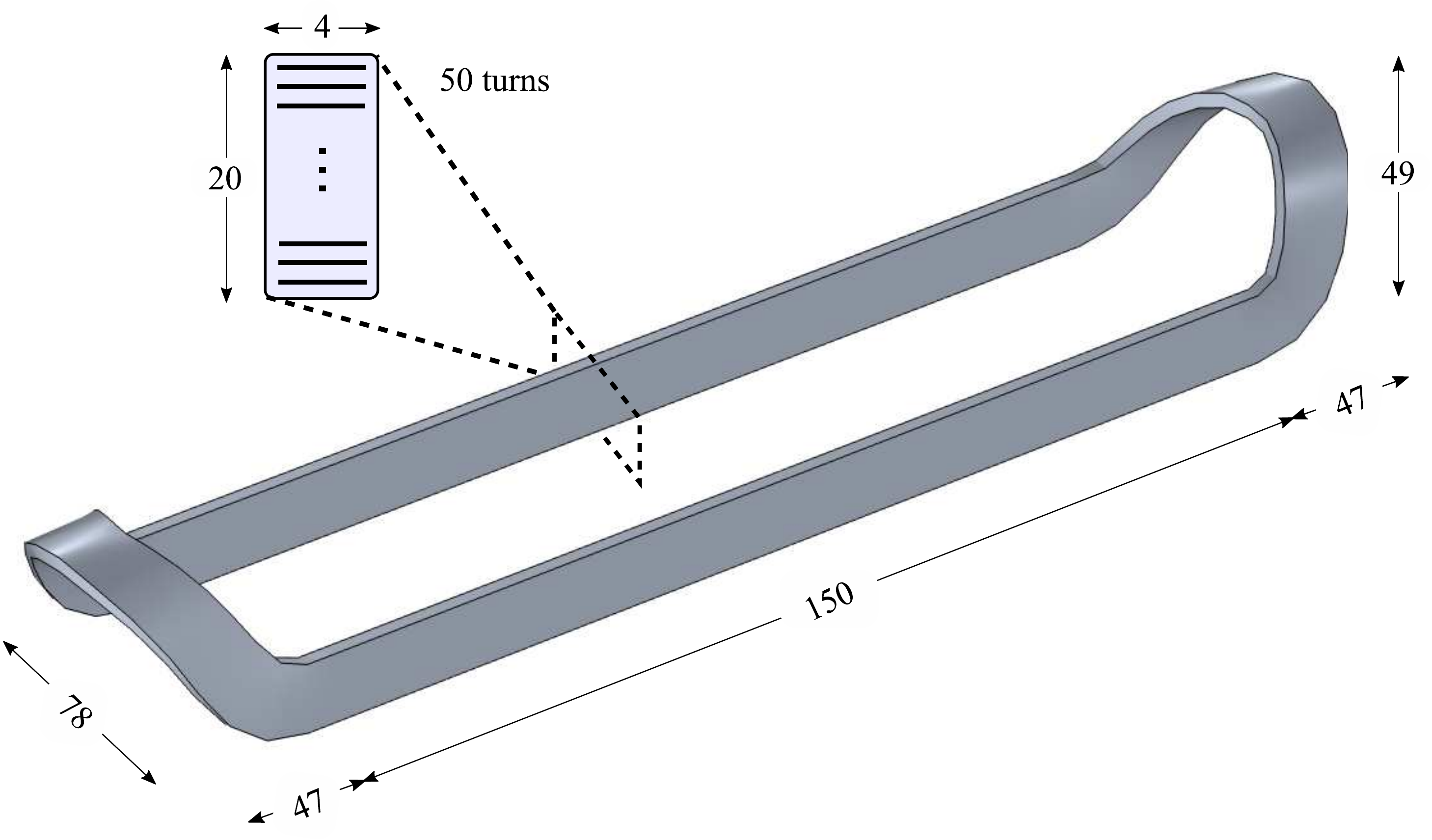} }
\caption{Geometry and dimensions of the saddle coil under study. The coil is made with 50 turns of HTS tape, which creates a stack \SI{4}{\milli\meter} wide and \SI{20}{\milli\meter} high in the cross-section of the coil. All the dimensions are in millimeters.}
\label{Saddle_geom}
\end{figure}

The second case of study is a saddle coil. This is a non-planar coil with two straight and two curved parts, which allow a \SI{90}{\degree} rotation of the cross-section between the middle of the circular and the straight region. The geometry and dimensions of the coil are presented in figure~\ref{Saddle_geom}. Only one-fourth of the coil was modeled by taking advantage of the symmetries. As it was done in the previous section, the critical current was estimated with a 2D model by following the procedure described in~\cite{zermeno_self-consistent_2015} and using the average criterion. The critical current of the coil is \SI{100}{\ampere}. This critical current was verified with a 3D  simulation where the current of the coil was ramped up (\SI{100}{\ampere\per\s}) until the critical current of the tape was reached. In this simulation, we saw at \SI{100}{\ampere} the full penetration of the coil with critical current density, and a change of the slope in the instantaneous losses which represents the operational limit established by the critical current.

This type of coil is used in the magnet sector and was proposed for the rotor and stator of superconducting electrical machines, where a proper inclination of the straight section can lead to a significant reduction in AC losses~\cite{rossi_hts_2021},~\cite{oomen_transposed-cable_2009},~\cite{vargas-llanos_influence_2021}. Two operating conditions are analyzed in this section, AC transport current with and without AC externally applied magnetic field.

\subsection{AC transport current without magnetic field}

\begin{figure}[hbt!]
\centerline{\includegraphics[width=0.45\textwidth]{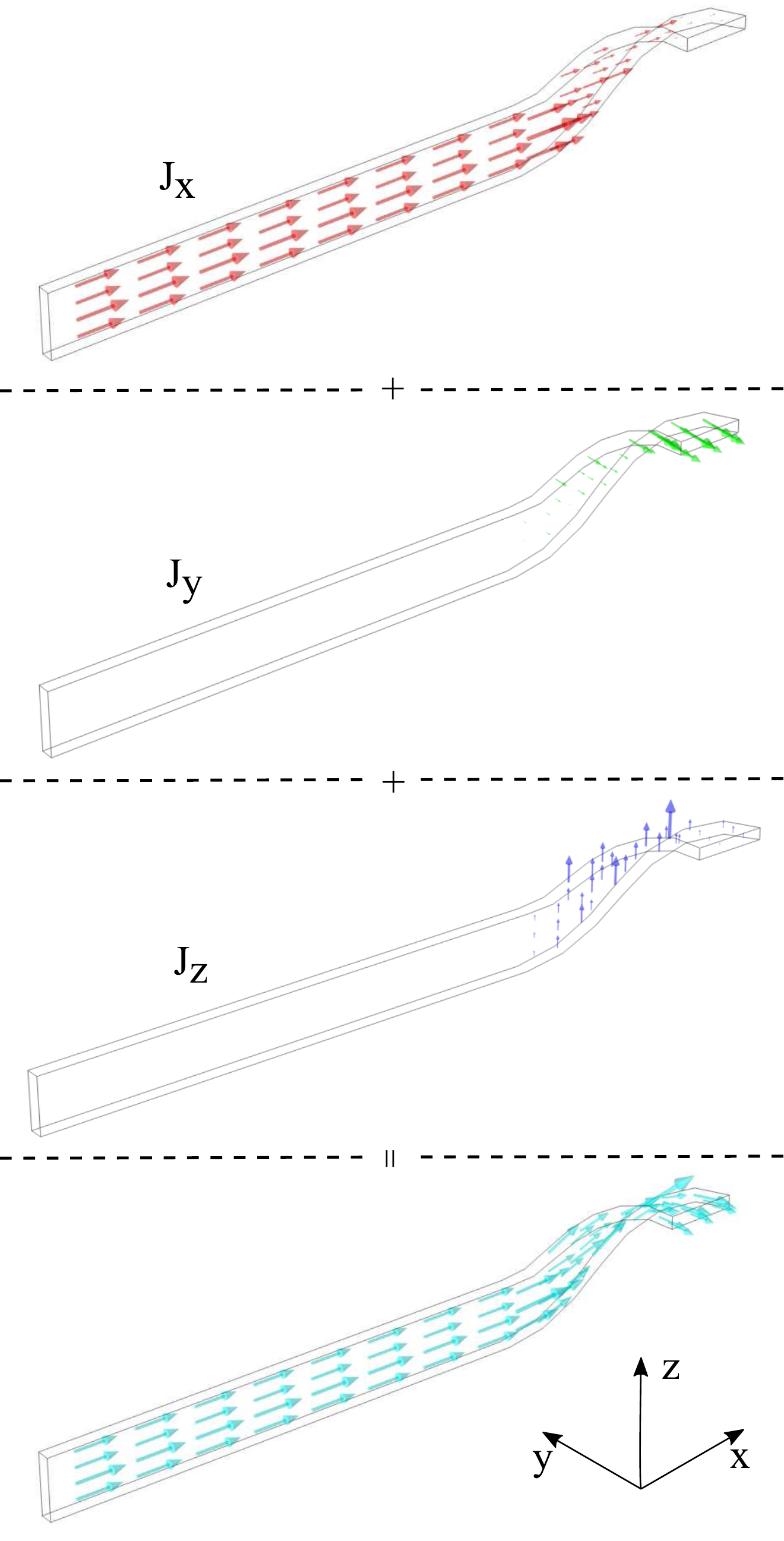} }
\caption{Distribution of the current components inside the saddle coil when the current reaches the maximum value. From top to bottom: x-component (${J}_x$), y-component (${J}_y$) and z-component (${J}_z$). The total current is depicted in light blue arrows at the bottom.}
\label{Saddle_current_arrows}
\end{figure}

The AC transport current (without external magnetic field) operating condition is easier to analyze for this complex geometry. Therefore, it was chosen as the first step in the analysis of the coil. Since this is a non-planar coil, the current vector has x-component, y-component and z-component ( figure~\ref{Saddle_current_arrows}). In the middle of the straight section, the current flows parallel to the x-axis. The curved section works as a transition zone for the current vector from the x-direction to the y-direction, by following the winding and position of the tapes. Due to the complex behavior of the current density ($\Vec{J}$) in 3D geometries, the cartesian components ($J_x, J_y, J_z$) are not suitable for representation purposes. A possible way to overcome this problem is to use the norm of $\Vec{J}$. However, we would not be able to appreciate the two fronts of currents in opposite directions during the AC cycle. Therefore, we use the dot product between the current density vector and the tangential vector parallel to the winding direction ($\Vec{e}_{t}$): 

\begin{equation} 
\frac{\Vec{J} \centerdot \Vec{e}_{t} }{J_\textrm{c}(B_{\parallel},B_{\bot})}.
\label{eq_J_Jc_3D_plot} 
\end{equation} 

The dot product between this unit vector, parallel to the longitudinal direction of the tape, and the current density keeps the same module and direction of $\Vec{J}$, by having a positive/negative value when the current density is in the same/opposite direction of $\Vec{e}_t$. The parallel and perpendicular (to the wide face of the tape) components of the magnetic field required for the local computation of $J_\textrm{c}(B_{\parallel},B_{\bot})$ can also be calculated in a more general expression with the dot product between the magnetic field and the other two unit (base) vectors associated to the definition of the curvilinear coordinate system:

\begin{equation} 
B_{\parallel} = \Vec{B} \centerdot \Vec{e}_{\parallel}
%\frac{\textbf{J} \centerdot \textbf{e}_{t} }{J_c(B_{\parallel},B_{\bot})}.
\label{eq_B_par} 
\end{equation} 

\begin{equation} 
B_{\bot} = \Vec{B} \centerdot \Vec{e}_{\bot}.
%\frac{\textbf{J} \centerdot \textbf{e}_{t} }{J_c(B_{\parallel},B_{\bot})}.
\label{eq_B_per} 
\end{equation} 

%We can appreciate in figure~\ref{Saddle_current_norm_3D} the current density behavior in the coil 

\begin{figure}[hbt!]
\centerline{\includegraphics[width=0.5\textwidth]{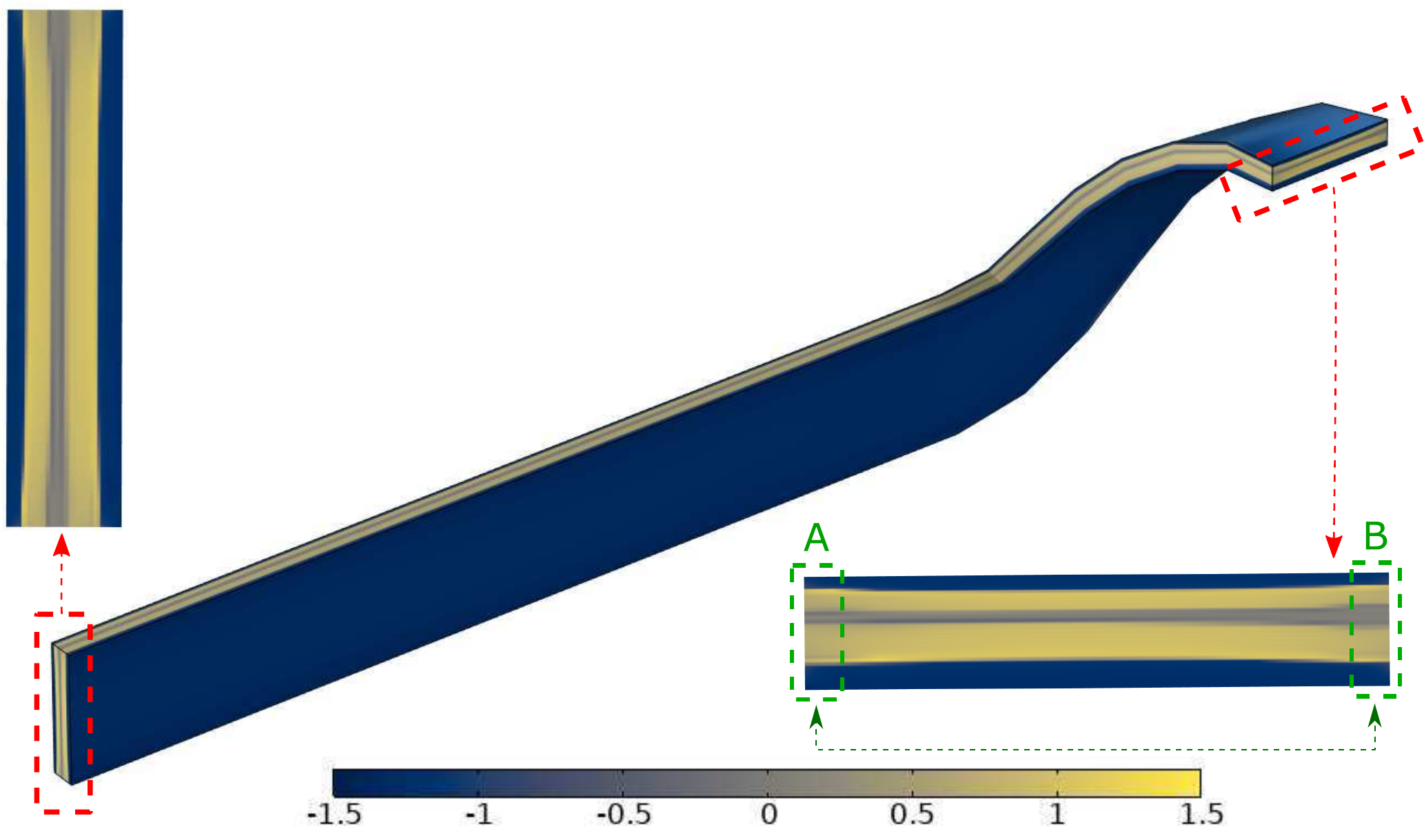} }
\caption{Normalized current density in the saddle coil computed with the 3D T-A homogenization model, for AC transport current ($f=\SI{50}{\hertz}$ and $I_{\rm peak}=\SI{100}{\ampere}$) when the current is equal to zero, and after the first half period of the sinusoidal cycle. The current density is normalized with the critical current density by using equation (\ref{eq_J_Jc_3D_plot}) to reflect the two fronts of current (positive and negative) in the direction tangential to the winding.}
\label{Saddle_current_norm_3D}
\end{figure}

In figure~\ref{Saddle_current_norm_3D} we represent the current density behavior in the coil for an AC transport current with a peak value of $\SI{100}{\ampere}$ and frequency $f=\SI{50}{\hertz}$, when the current is equal to zero, and after the first half period of the sinusoidal cycle. The current density penetration and behavior is very similar in the middle of the straight and circular sections. There is only a very small difference between the inner and the outer part of the curved section, indicated with green arrows as A and B in figure~\ref{Saddle_current_norm_3D}. This small difference is an effect of the self-field of the coil, which tends to be higher inside the coil than outside. According to these results, a 2D model that represents the middle of the straight section of the saddle coil can provide a good approximation of the current density behavior and a first estimation of the AC losses. This bi-dimensional model is simpler, faster, and easier to implement; and is commonly used to have first estimations of important parameters such as AC losses and magnetic field. Therefore, we decided to build the 2D model to verify whether this assumption is valid for the proposed case study.

\begin{figure}[hbt!]
\centerline{\includegraphics[width=0.5\textwidth]{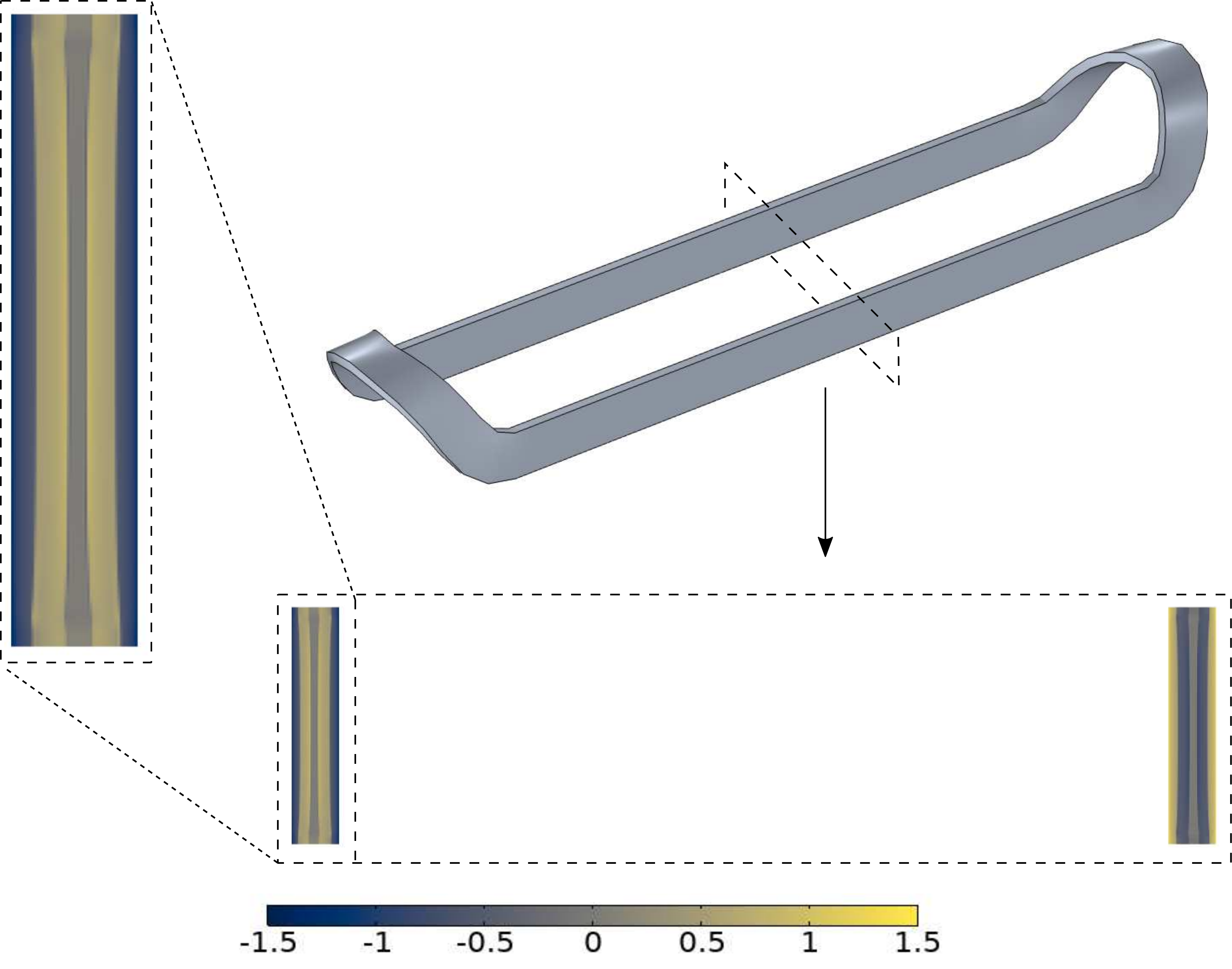} }
\caption{Normalized current density in the middle of the straight section of the saddle coil computed with the 2D T-A homogenization model, for AC transport current ($f=\SI{50}{\hertz}$ and $I_{\rm peak}=\SI{100}{\ampere}$) when the current is equal to zero, and after the first half period of the sinusoidal cycle.}
\label{Saddle_current_norm_2D}
\end{figure}

Figure~\ref{Saddle_current_norm_2D} shows the current density behavior obtained with the 2D model that represents the middle of the straight section of the coil, for the same operating point and condition used in figure~\ref{Saddle_current_norm_3D}. The current density behavior is similar between the 2D and 3D models for the straight part of the coil. In the curved section, there is a small difference. As mentioned before, this difference is related to the self-field in 3D, which is a result of the interaction of the different parts (straight and curved) of the coil and can not be taken into account by a 2D model.     

\begin{figure}[hbt!]
\centerline{\includegraphics[width=0.5\textwidth]{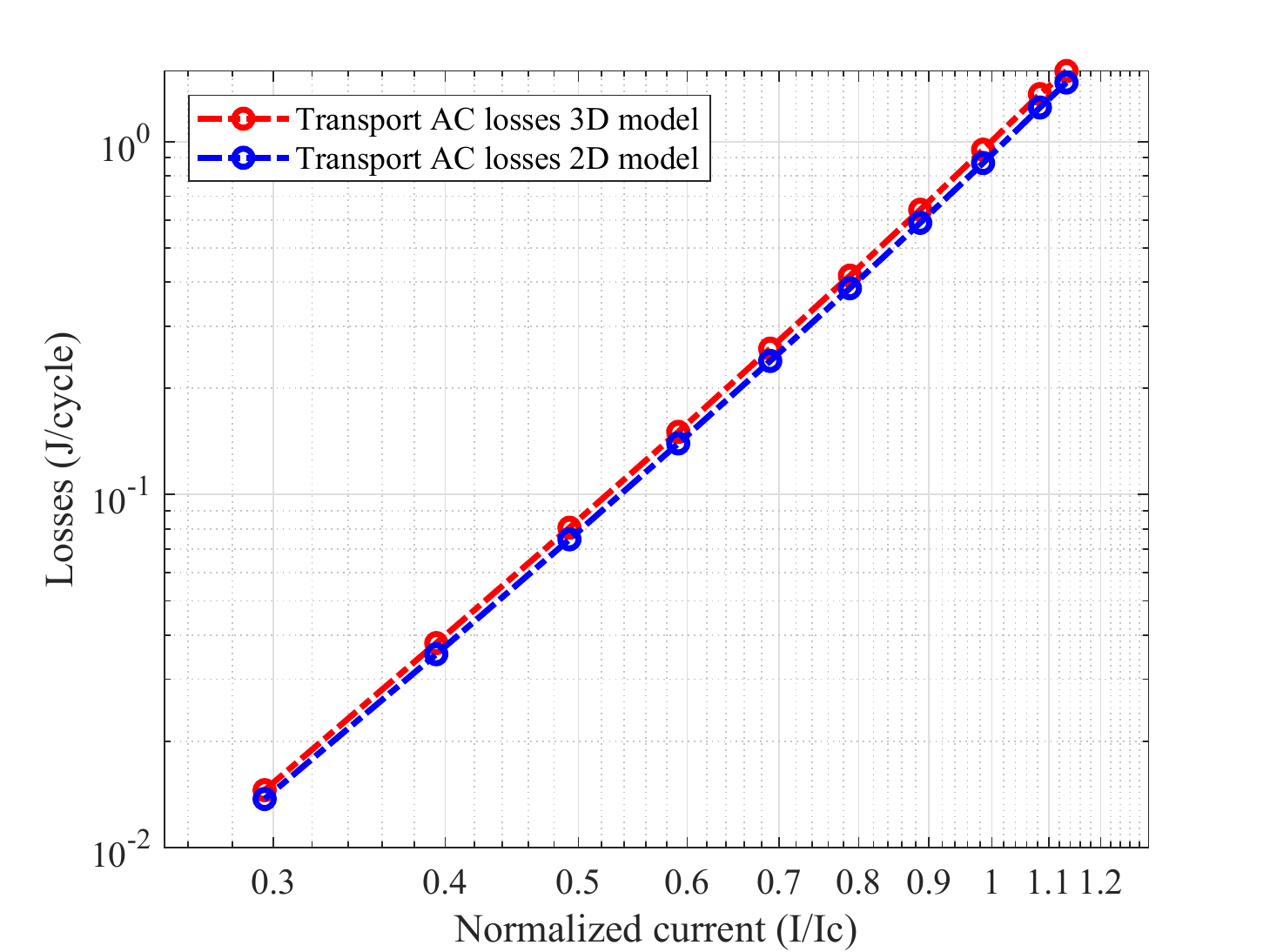} }
\caption{AC transport losses comparison between the 3D T-A homogenization and the 2D T-A homogenization model that considers the cross-section of the middle of the straight part of the coil.}
\label{Saddle_current_AC_transport_losses}
\end{figure}

Figure~\ref{Saddle_current_AC_transport_losses} presents the behavior of the AC transport losses calculated with the T-A 3D and 2D homogenized models. The losses estimated with both models are in good agreement for all the range of current amplitudes under analysis, with a maximum relative difference of \SI{8.4}{\percent}. The small differences noticed in the current density in the curved section do not have a meaningful impact on the AC transport losses. These differences vanish in the behavior of the whole coil, where the straight part behavior is dominant in the AC transport losses. These results confirm the aforementioned hypothesis. The 2D model can be used as a first estimation of important parameters such as AC transport current losses for this specific case. However, the magnetic field around the coils in most of the practical applications is a combination of self-field and interaction with the environment. This yields a more complex operating condition that a 2D model can not fully predict. 

\subsection{AC Transport current and AC external magnetic field}

In this sub-section, we analyze the behavior of the saddle coil in the case of AC transport current (with a peak value of $\SI{100}{\ampere}$ and frequency $f=\SI{50}{\hertz}$) and AC external magnetic field. The external magnetic field is uniform, sinusoidal (\SI{100}{\milli\tesla} in magnitude and frequency $f=\SI{50}{\hertz}$) and applied in the z-direction, in phase with the magnetic field produced by the coil. This can be seen as the magnetic field produced by the rotor of an electrical machine that reaches one of the stator superconducting coils. However, the magnetic field in the stator of a superconducting electrical machine is more complex than the presented operating condition. In an electrical machine, the magnetic field changes around the stator and is not necessary sinusoidal in time (it depends on the properties of the materials, the machine and coil design...)~\cite{vargas-llanos_influence_2021}. Therefore, the case presented in this subsection is a simplified version of a more complex operating condition, which allows us to study the behavior of the coil with AC transport current and AC external magnetic field. The analysis is not intended to predict the behavior of the coil in an electrical machine environment. 

%%%%%%%%%%%%%%%%%%%%%%%%%%%%%%%%%%%%%%%%%%%%
\begin{figure*}[hbt!]
  \centering
  \includegraphics[width=0.8\textwidth]{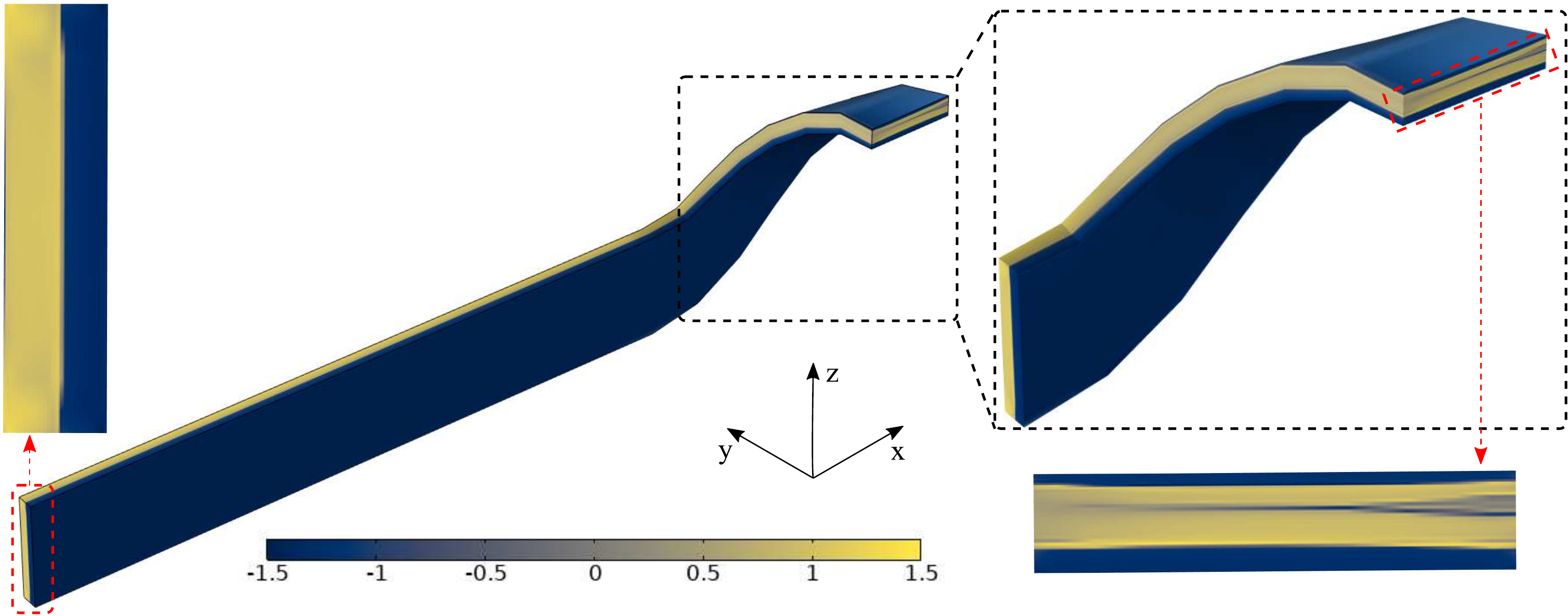}
  \caption{Normalized current density in one-fourth of the saddle coil for AC transport current ($f=\SI{50}{\hertz}$ and $I_{\rm peak}=\SI{100}{\ampere}$) and external magnetic field ($f=\SI{50}{\hertz}$ and $B_{\rm ext}=\SI{100}{\milli\tesla}$) in $z$-direction and in phase with the coil self-field, when the current is equal to zero and after the first half period of the sinusoidal cycle. A Zoom in the curved section of the coil is presented on the right to show the transition between a full and partial penetrated area.}
  \label{Saddle_coil_current_and_Magn_field}
\end{figure*}
%%%%%%%%%%%%%%%%%%%%%%%%%%%%%%%%%%%%%%%%%%%%%

Since the external magnetic field is in the z-direction and in phase with the self-field of the coil, we can expect a higher magnetic field magnitude in the inner part (where the external and self field are in the same direction) and lower in the outer part of the coil (where the external and self field are in opposite direction). Moreover, the external magnetic field will be perpendicular to the tape in the straight section and parallel in the curved one. This will create a more complex behavior than in the transport current operating condition. 

Figure~\ref{Saddle_coil_current_and_Magn_field} shows the current density behavior when the transport current is equal to zero, and after the first half period of the sinusoidal cycle. As it can be observed, the straight section is fully penetrated with critical current density. We can notice one front of current coming from the inner side of the homogenized stack, and a very thin current front in the outer part of the middle of the straight section (where the external and self field are in opposite directions). The curved section shows a transit behavior in space from a fully penetrated zone (straight part) to a partially penetrated one. In the middle, we can notice two fronts of current penetrating the stack at the top and bottom. The inner part is fully penetrated, but the outer part has gray areas that show a partial penetration of the current in the coil. In comparison with the transport current operating condition (figure~\ref{Saddle_current_norm_3D}), this behavior indicates that the external magnetic field has a stronger influence in the straight section of the coil than in the curved one, due to the torsion of the stack in the geometry of the coil from the straight part (where the external magnetic field is perpendicular to the tape) into the curved part (where the external magnetic field is parallel to the tape).

The total AC losses in the coil for this operating condition are \SI{3.62}{\joule} per cycle, which represents more than three times the AC losses of the transport current case with the same current amplitude (\SI{0.95}{\joule} per cycle). The complex current density behavior and the big increase in the AC losses observed in this sub-section suggest that a 3D approach is better suited for the analysis of this operating condition. This 3D approach provides a more accurate estimation of losses and allows an electromagnetic analysis that can not be done with a 2D approximation.

\section{D-Shape coil model and analysis}\label{Section_D-Shape coil}

\begin{figure}[htb!]
\centerline{\includegraphics[width=0.35\textwidth]{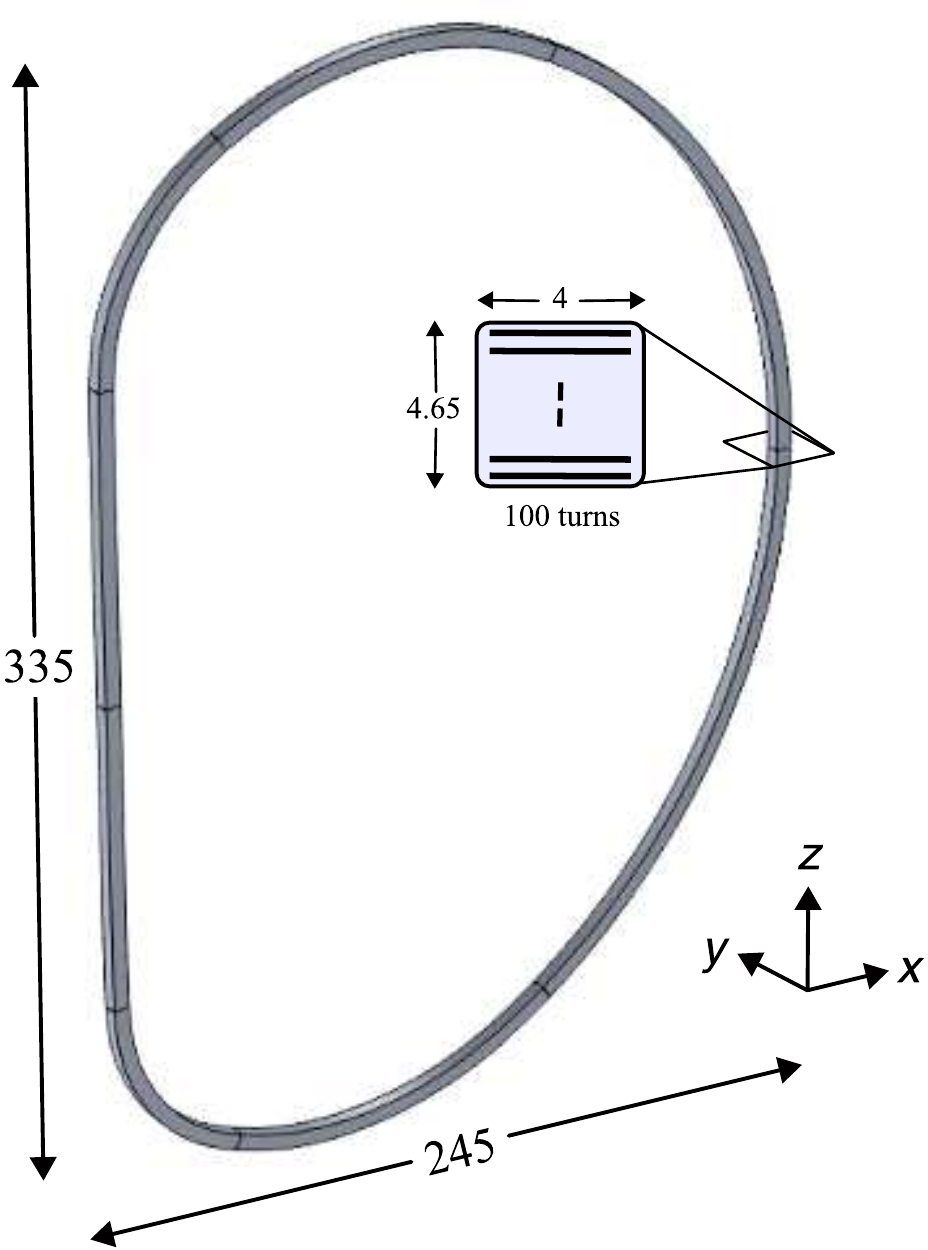} }
\caption{Geometry of the D-shaped coil with the design parameters taken from~\cite{liu_electromagnetic_2014}. The HTS conductor cross-section is used for the case study. All the dimensions are in millimetres.}
\label{FH_D-shaped_coil_Geometry}
\end{figure}

In this section, the geometry of a D-shaped coil, used in Tokamak fusion devices, is investigated~\cite{savary_toroidal_2012},~\cite{liu_progress_2020}. The shape and dimensions of the coil are shown in figure~\ref{FH_D-shaped_coil_Geometry} (design parameters presented in~\cite{liu_electromagnetic_2014}). The geometrical design of this type of coil features a straight inner leg and curved segments with different radii. This geometry is reduced to a quarter model by considering symmetry with the horizontal and vertical planes at the center of the coil.

%%%%%%%%%%%%%%%%%%%%%%%%%%%%%%%%%%%%%%%%%%%%
\begin{figure*}[htb!]
  \centering
  \includegraphics[width=0.70\textwidth]{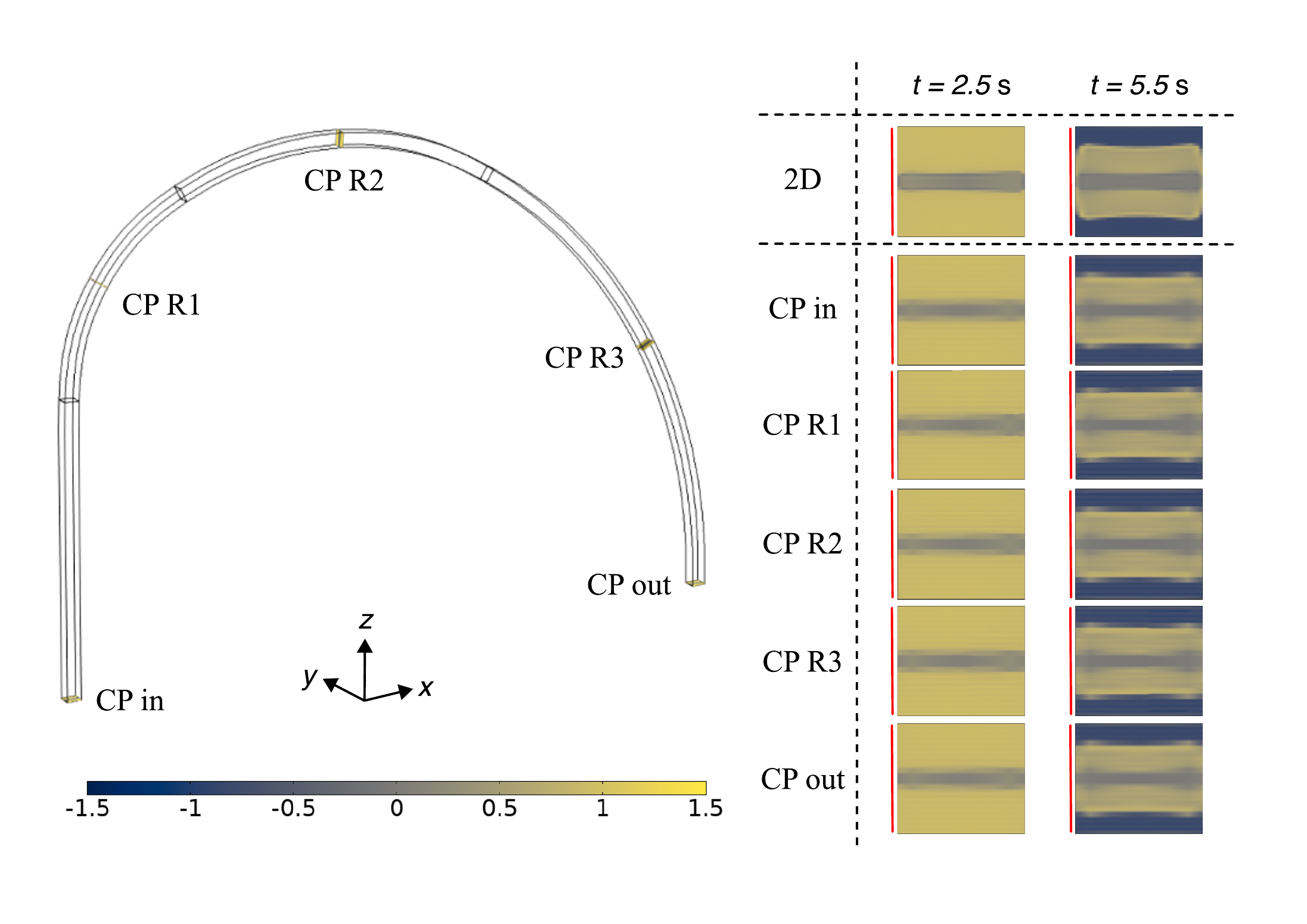}
  \caption{Normalized current density at the indicated cut planes (CP) of the 3D model in comparison with the simplified 2D model. The red lines indicate the inside of the coil. The current penetration is shown at $t = \SI{2.5}{\second}$ (during the relaxation time after the ramp-up phase) at a constant current of \SI{50}{\ampere}, and at $t = \SI{5.5}{\second}$ (during the relaxation time after the ramp-down phase) at a constant current of \SI{0}{\ampere}.}
  \label{FH_D-shaped_coil_current_penetration}
\end{figure*}
%%%%%%%%%%%%%%%%%%%%%%%%%%%%%%%%%%%%%%%%%%%%%

\begin{figure}[htb!]
\centerline{\includegraphics[width=0.5\textwidth]{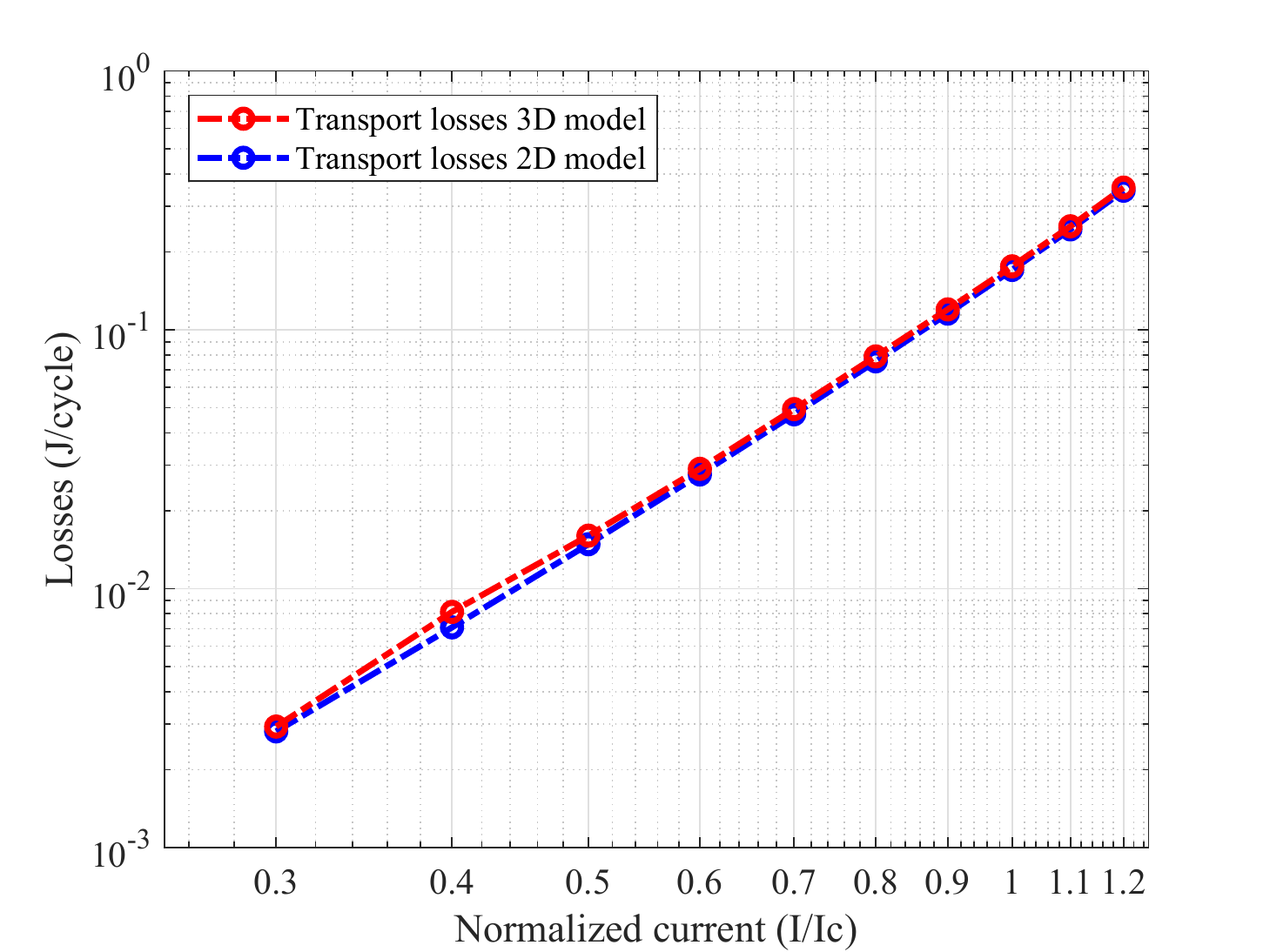} }
\caption{Transport losses for the D-shaped coil model in 3D compared to the simplified axial symmetric 2D model. The 2D model uses the mean radius of the D-shaped coil.}
\label{FH_D-shaped_coil_validation}
\end{figure}

\begin{figure}[htb!]
\centerline{\includegraphics[width=0.5\textwidth]{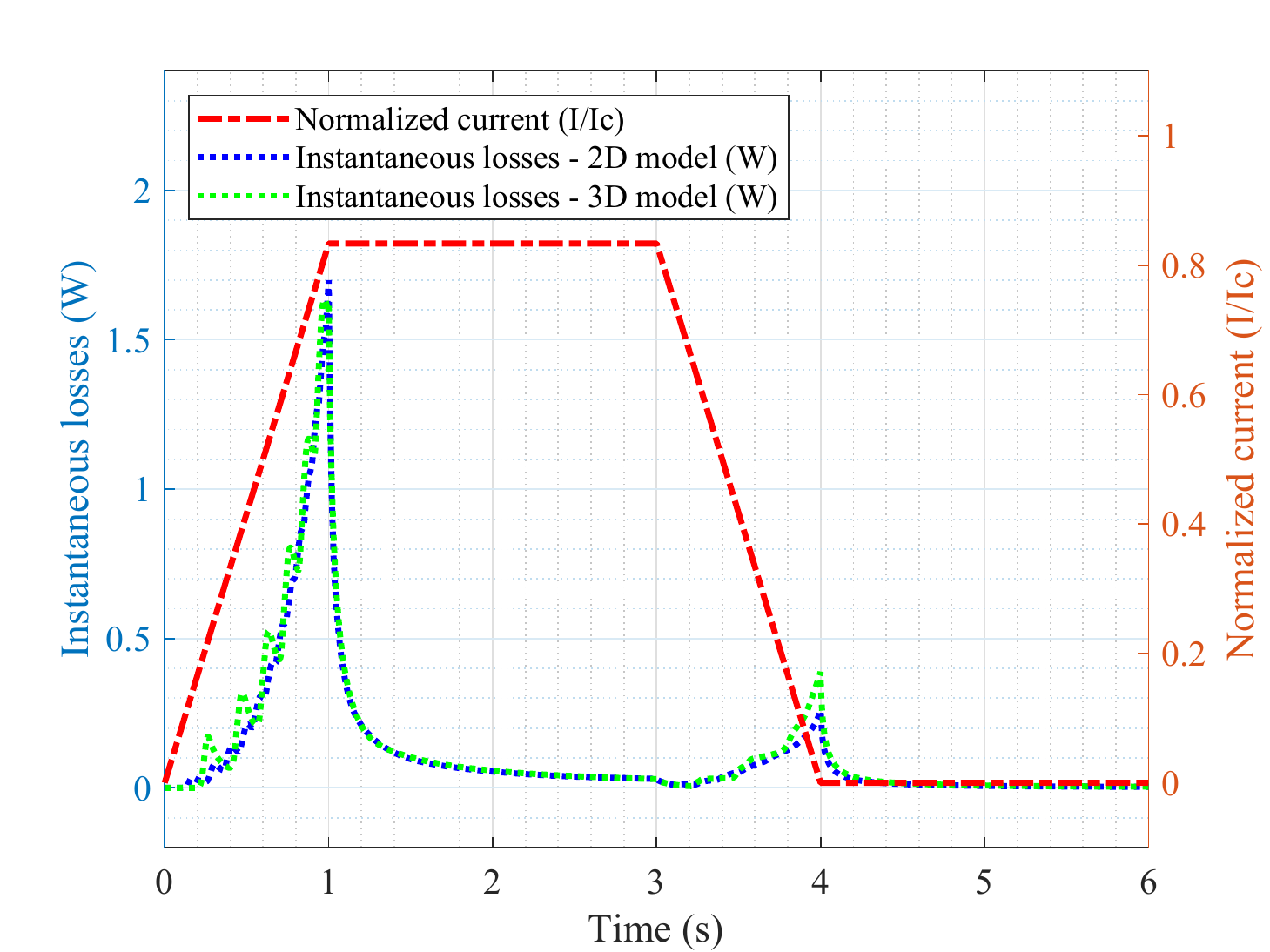} }
\caption{Current ramp-up/down process (load cycle) and instantaneous losses in the D-shaped coil model computed with the 3D and 2D T-A homogenization models.}
\label{FH_D-shaped_coil_load_cycle}
\end{figure}

Due to the high ratio of the coil dimension to the HTS conductor cross-section, the model has a large number of degrees of freedom. To keep the simulation within a manageable timeframe, the meshing of the geometry needs to be carefully chosen. It is important to use as few elements as possible while not compromising the accuracy of the model. This limits the number of elements that can be used to mesh the HTS conductor cross-section. By using a structured, quadrilateral mesh in the superconducting domain (10 elements equally spaced along the tape width and 8 elements along the width of the stack) and a free tetrahedral mesh in the surrounding air domain, the whole geometry can be meshed with 22822 elements in total, resulting in 293477 degrees of freedom. %({\FG maybe give some info about the mesh?})

In the first step, the modeling approach with the D-shaped geometry is validated by using the same HTS cross-section presented in the previous section. For the validation, the transport losses of the 3D model are compared to a 2D axial symmetric model. The mean radius is calculated from the height and width of the 3D geometry ($R_m = \SI{145}{\milli\meter}$) to approximate the D-shaped coil in 2D. The results for the comparison of transport currents ranging from \SI{30}{\ampere} to \SI{120}{\ampere} are shown in figure~\ref{FH_D-shaped_coil_validation}. It can be seen that the agreement between the models is very good, with a maximum relative difference of \SI{12.9}{\percent}.

Subsequently, a study case for a load cycle is evaluated by using the HTS cross-section shown in figure~\ref{FH_D-shaped_coil_Geometry}. The critical current for this coil is \SI{60}{\ampere}. In the first second of the load cycle, the transport current is ramped up from \SI{0}{\ampere} to \SI{50}{\ampere}, followed by a \SI{2}{\second} relaxation time. Then the current is ramped down over \SI{1}{\second} from \SI{50}{\ampere} to \SI{0}{\ampere}, and at the end another \SI{2}{\second} are given as relaxation time. This load cycle is shown in figure~\ref{FH_D-shaped_coil_load_cycle} by the red dashed line. The blue and green dashed lines correspond to the power dissipation calculated with the 2D, and 3D models respectively. 

There are two distinct peaks in the losses, which occur at the end of each ramp phase (up/down). During the ramp-up of the current, the losses of the 3D model show a ripple effect, which can be attributed to the mesh distribution of the conductor area. This behavior is amplified at lower currents, if we keep the same mesh, and it can also be observed in coarsely meshed 2D models.

The instantaneous losses at the end of the ramp-up phase at \SI{1}{\second} are \SI{1.69}{\watt} for the 2D model and \SI{1.61}{\watt} for the 3D model. For the second peak at the end of the ramp-down phase at \SI{4}{\second}, the instantaneous losses are \SI{0.26}{\watt} and \SI{0.39}{\watt}, for 2D and 3D respectively. The evaluation of the losses for the whole load cycle shows that there is a discrepancy, and the 2D model underestimates the losses for this complex geometry by \SI{11.54}{\percent}. The losses amount to \SI{0.69}{\joule} and \SI{0.78}{\joule} for the whole load cycle, for the 2D and 3D models respectively. In figure~\ref{FH_D-shaped_coil_current_penetration}, the current penetration is shown at \SI{2.5}{\second} (during the stabilization after the initial ramp phase) and at \SI{5.5}{\second} (after the ramp down). On the left side of the figure, the locations of the cut planes are indicated in the geometry. The current distribution is shown in the respective cut planes of the 3D model and compared to the solution from the 2D axial symmetric model. In the cut planes \emph{CP in} and \emph{CP out}, the distribution is almost identical to the 2D model. The largest difference can be observed in the cut plane of the segment with the smallest radius (\emph{CP R1}). The current penetration at the inside of the coil is higher than at the outside of the coil. This can be explained by the influence of the self-field, which is larger in the inner part of the segment with the highest curvature. The 2D and 3D models show a good agreement. The computation time for a \SI{1}{\second} ramp phase is about \SI{20}{\hour} for the 3D model, while the 2D model finishes within minutes. For a detailed investigation of a D-shaped coil, the 3D model can provide a better understanding of the electromagnetic behavior.

\section{Twisted coil model and analysis}\label{Section_Twisted coil}
%Critical current = 60 A - Methodology: The current was increased (ramp 100 A/s) and the change of the slop in the AC losses was detected at 60 A/saturation of the coil's cross-section.

We study in this section a more complex geometry that has no straight parts. The twisted coil has only curved sections with a continuous twist along the length of the coil. The model and development of coils with this kind of geometry can be beneficial for the fusion sector~\cite{rummel_commissioning_2017},~\cite{almagri_helically_1999}. However, the case under analysis is not part of any specific application. As mentioned in section~\ref{Section_Saddle coil}, the detailed geometry and electromagnetic environment are necessary for any analysis of superconducting devices and applications, since they can strongly influence the losses and electromagnetic behavior of the coils. Therefore, this case of study can be seen as a complete curved and twisted coil analysis that shows the potential application of the modeling approach in a coil similar to the one used in the fusion sector.       

\begin{figure}[htb!]
\centerline{\includegraphics[width=0.4\textwidth]{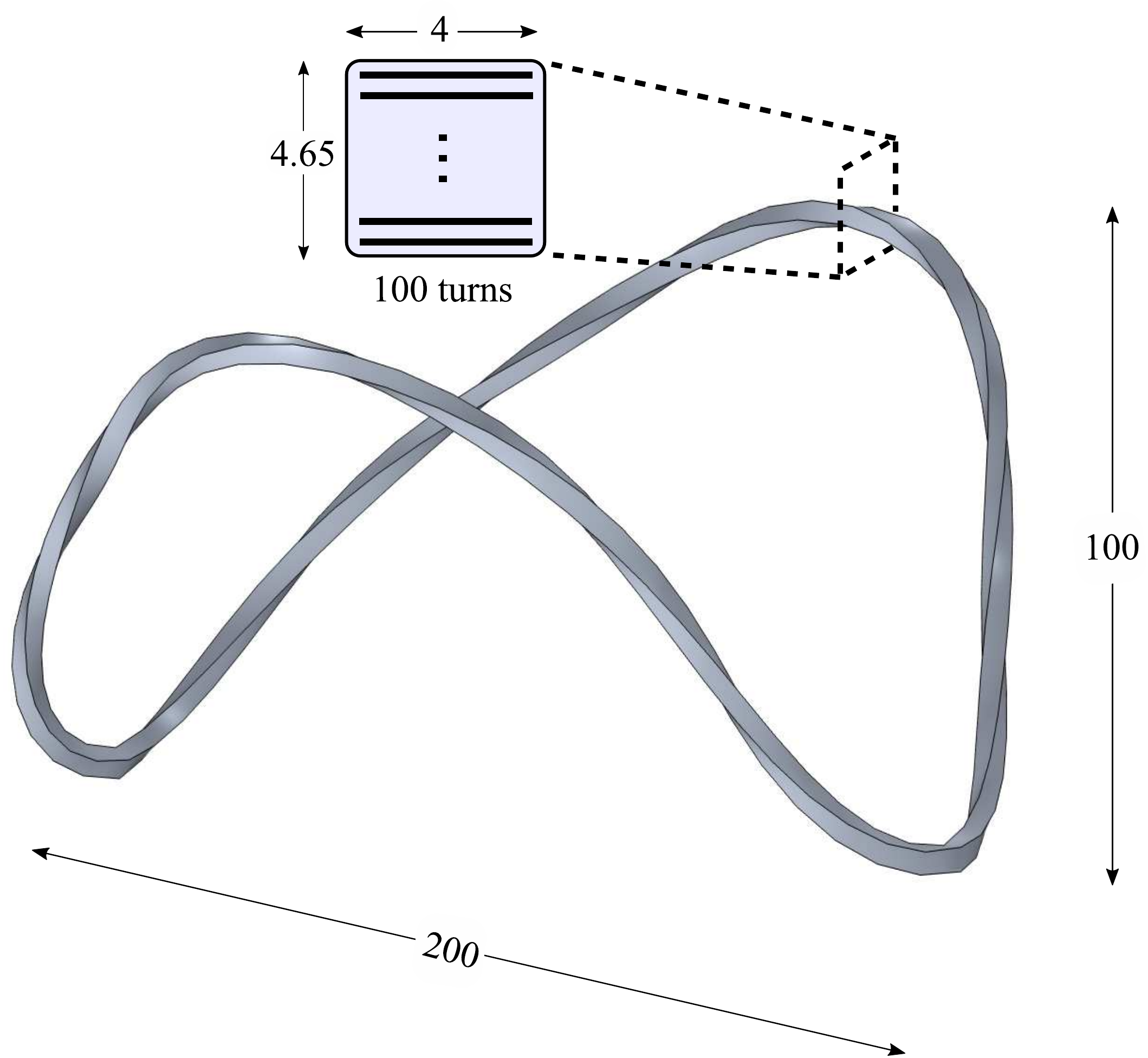} }
\caption{Geometry and dimensions of the twisted coil under study. The coil is made with 100 turns of HTS tape, which creates a stack \SI{4}{\milli\meter} wide and \SI{4.65}{\milli\meter} height in the cross-section of the coil. All the dimensions are in millimeters.}
\label{Twisted_geom}
\end{figure}

The geometry and dimensions of the coil are presented in figure~\ref{Twisted_geom}. As it can be {noticed}, it has one plane of symmetry. Therefore, only one-half of the coil was modeled. The critical current of this coil is \SI{60}{\ampere}. 

\begin{figure}[hbt!]
\centerline{\includegraphics[width=0.5\textwidth]{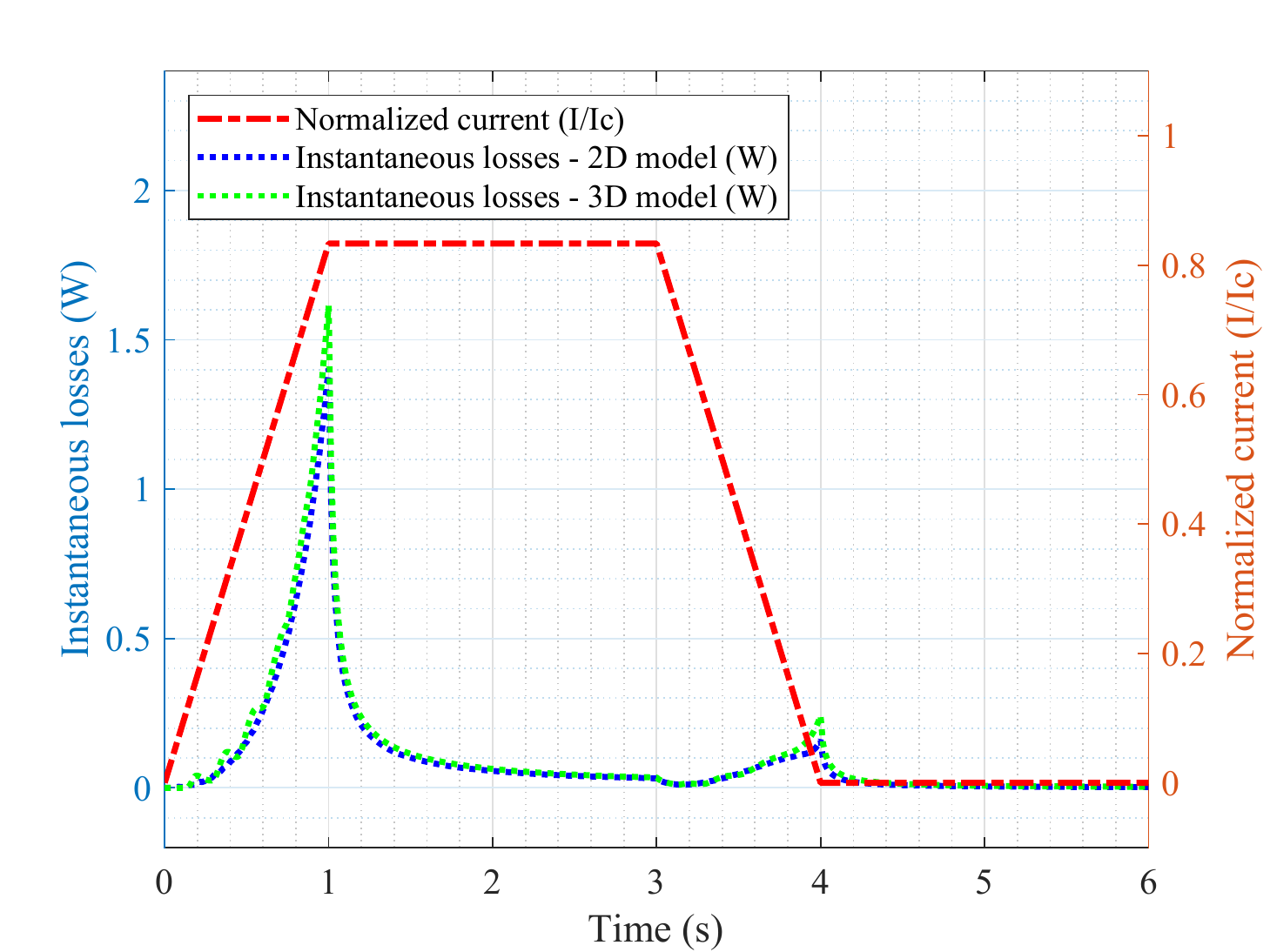} }
\caption{Current ramp-up/down process (load cycle) and instantaneous losses behavior in the twisted coil computed with the 3D and 2D T-A homogenization models.}
\label{Twisted_coil_losses}
\end{figure}

%%%%%%%%%%%%%%%%%%%%%%%%%%%%%%%%%%%%%%%%%%%%
\begin{figure*}[htb!]
  \centering
  \includegraphics[width=0.8\textwidth]{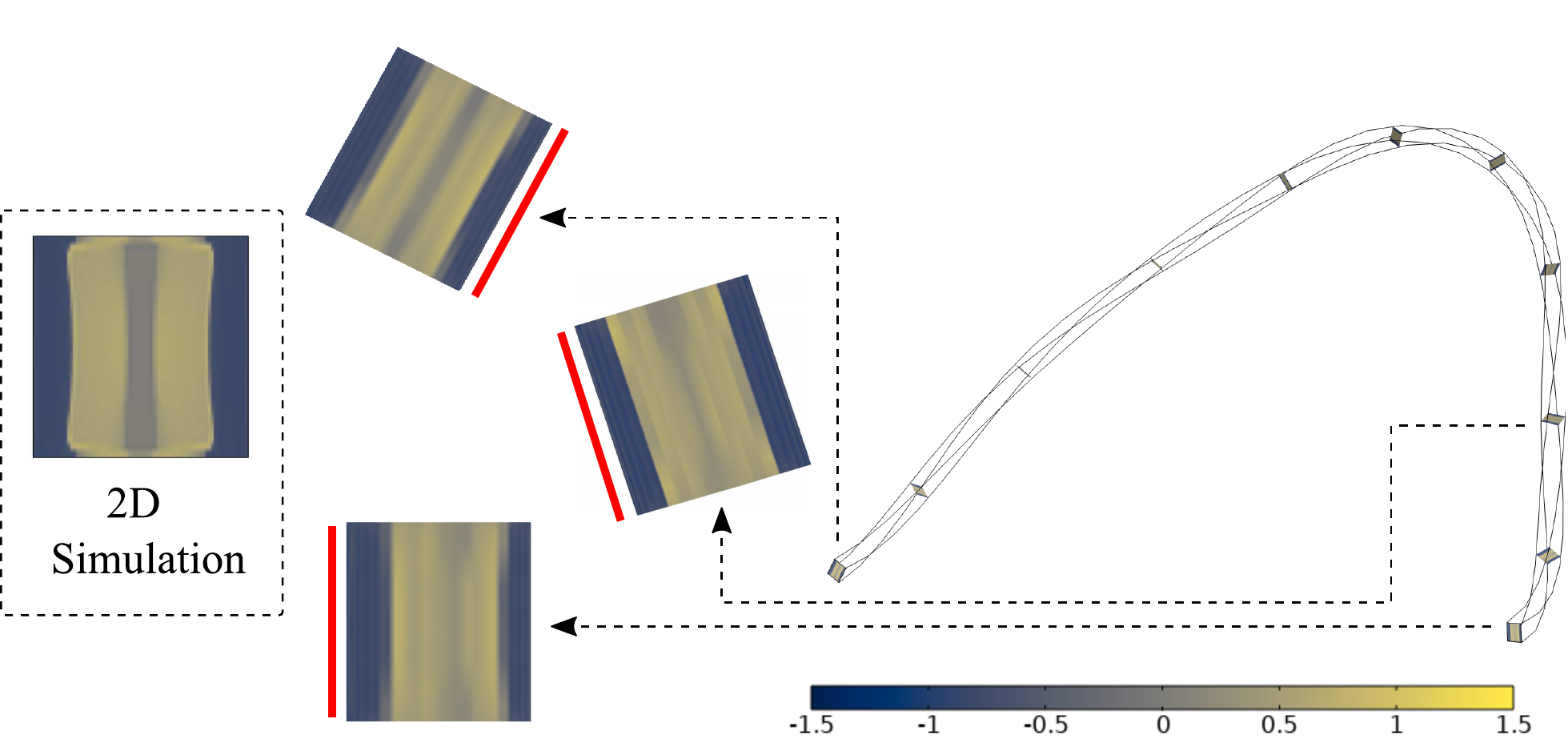}
  \caption{Normalized current density in one half of the twisted coil at $t=\SI{5.5}{\s}$ (during the relaxation time after the ramp-down process). A zoom in the cross-section of the coil at three different locations is presented on the left together with the behavior of the normalized current density in the 2D model at the same instant in time.}
  \label{Twisted_coil_current_penetration}
\end{figure*}
%%%%%%%%%%%%%%%%%%%%%%%%%%%%%%%%%%%%%%%%%%%%%

We considered a DC transport current as the operating condition of the coil. For this reason, we simulate the current ramp-up/down process as one possible test or operating cycle. As it was done in the previous case study, we first ramp-up the current from \SI{0}{\ampere} to \SI{50}{\ampere} (\SI{83.33}{\percent} of the critical current) in one second and wait \SI{2}{\s} as relaxation time. Then, we ramp-down the current from \SI{50}{\ampere} to \SI{0}{\ampere} in one second and wait \SI{2}{\s} as relaxation time. The overall load cycle and losses behavior can be observed in figure~\ref{Twisted_coil_losses}. We have two peaks in the losses of the coil related to the ramp-up/down periods of the current. These losses were estimated by using the aforementioned 3D approach and a simplified 2D T-A homogenized model of the cross-section of the coil. The 3D model estimates a peak in the instantaneous losses due to the first ramp of \SI{1.61}{\watt} and the 2D model estimates a peak of \SI{1.40}{\watt}. This difference can be related to the fact that the 2D model can not consider the continuous twist and curves along the length of the coil. Similarly, the 3D model also estimates a slightly higher value for the second peak in the instantaneous losses due to the second ramp (\SI{0.24}{\watt}), in comparison with the 2D model (\SI{0.16}{\watt}).

Figure~\ref{Twisted_coil_current_penetration} shows the behavior of the current density in half of the coil at t=\SI{5.5}{\s} (during the relaxation time after the ramp-down). In this figure, the red line highlights the position of the same side of the coil cross-section along the twist. The current penetration is sometimes higher on one side of the homogenized stack, sometimes higher on the other side, and in some cases looks symmetrical. This behavior is related to the interaction of the current density with the self-field produced by different parts of the curved and twisted geometry, which can cause local higher penetration and saturation of current. This behavior can not be reproduced by a 2D model. Figure~\ref{Twisted_coil_current_penetration} also shows the current penetration in the 2D T-A homogenized model for the same time reference. As it can be observed, the current penetration is more symmetrical in the 2D model. This can cause the discrepancies in the estimation of losses presented before. These discrepancies can be higher depending on the geometry, local effects, and operating conditions.     

In this case of study, the 2D T-A homogenized model can provide a first estimation of the losses due to the current ramp-up/down process (without external magnetic field). However, a 3D model can provide a better electromagnetic analysis of local effects (current saturation and high penetration zones) that can cause higher losses at different operating conditions or electromagnetic environments.

\section{Conclusion}\label{Conclu}

The 3D T-A homogenization modeling technique was used to study the behavior of superconducting coils with complex 3D geometries under different operating conditions. The proposed normal vector approach based on a curvilinear coordinate system definition allowed an easier implementation (in comparison with the currently available 3D H homogenization) despite the complexity of the geometry. Therefore, it represents a practical tool for the analysis of HTS devices with complex shapes.

We validated first the modeling approach with the racetrack coil studied in~\cite{zermeno_3d_2014} by comparing the results with the 3D H homogenization technique. Then, we modeled and analyzed a saddle coil, which represented a case study with potential applications in the magnet and superconducting electrical machine sectors. The operating conditions under study were AC transport current and AC transport current with AC externally applied magnetic field. We compared the results with a 2D T-A homogenized model, which gave a good approximation of the losses for the first operating condition, but can not be used for the electromagnetic analysis of the second one.  

A D-Shape coil and a twisted coil (without straight parts) were also modeled and analyzed. These cases have potential applications in the fusion energy sector. The operating condition considered for these coils was DC transport current. For this reason, the ramp-up/down process was simulated.

For the coils simulated in this publication, the losses due to AC or the ramp-up/down process can be estimated with a 2D model. This approach provides a good approximation and offers a great reduction of degrees of freedom due to the reduction of dimension. Therefore, it represents a good compromise between accuracy and computation time. However, the behaviour of the coils due to transport current and external magnetic field can be more complex. For this operating condition, the 3D model can provide a better estimation of losses and detect zones with local saturation of current, as was seen in the saddle coil simulations. The operating conditions of HTS coils in the majority of applications (such as electrical machines and fusion experiments) involve transport current and magnetic field. In these cases, the study of the interaction with other coils, magnets, and materials in intricate 3D geometries is only feasible with a 3D model.

These analyzes show the versatility of the modeling approach and its efficiency for AC loss estimation in coils. Moreover, the studies support and encourage the modeling and development of HTS coils with complex geometries, necessary to overcome the technological challenges of superconducting devices in multiple fields and applications. 

\section*{Acknowledgment}
This work was partly supported by the EU-funded Hi-SCALE COST Action CA19108.

\bibliographystyle{IEEEtran}
%\bibliography{My_Library_30112021.bib}
%\bibliography{My_Library_13012022.bib}
\bibliography{My_Library_19052022.bib}

\end{document}